\documentclass[reprint,
superscriptaddress,
 amsmath,amssymb,
 aps,
pra,
]{revtex4-1}
\usepackage{dsfont}
\newcommand{\id}{\mathds{1}}
\usepackage{nicefrac}
\usepackage{braket}
\usepackage{xcolor}
\usepackage{graphicx}
\usepackage{dcolumn}
\usepackage{bm}
\usepackage{hyperref}
\usepackage[caption=false]{subfig}
\usepackage[normalem]{ulem}

\usepackage{algorithm}
\usepackage{algpseudocode}
\usepackage{algorithmicx}

\usepackage[hang,flushmargin]{footmisc}

\begin{document}

\preprint{APS/123-QED}

\title{Measurement device-independent quantum key distribution with passive, time-dependent source side-channels}

\author{J. Eli Bourassa}
\thanks{These two authors contributed equally. \newline eli.bourassa@mail.utoronto.ca \newline amita.gnanapandithan@mail.utoronto.ca}
\affiliation{Department of Physics, University of Toronto, Toronto, Canada}

\author{Amita Gnanapandithan}
\thanks{These two authors contributed equally. \newline eli.bourassa@mail.utoronto.ca \newline amita.gnanapandithan@mail.utoronto.ca}
\affiliation{Department of Electrical and Computer Engineering, University of Toronto, Toronto, Canada}

\author{Li Qian}
\affiliation{Department of Electrical and Computer Engineering, University of Toronto, Toronto, Canada}
\affiliation{Center for Quantum Information and Quantum Control, University of Toronto, Toronto, Canada}

\author{Hoi-Kwong Lo}
\affiliation{Department of Physics, University of Toronto, Toronto, Canada}
\affiliation{Department of Electrical and Computer Engineering, University of Toronto, Toronto, Canada}
\affiliation{Center for Quantum Information and Quantum Control, University of Toronto, Toronto, Canada}
\affiliation{Department of Physics, University of Hong Kong, Hong Kong}

\date{\today}

\begin{abstract}
While measurement-device-independent (MDI) quantum key distribution (QKD) allows two trusted parties to establish a shared secret key from a distance without needing to trust a central detection node, their quantum sources must be well-characterized, with side-channels at the source posing the greatest loophole to the protocol's security. In this paper, we identify a time-dependent side-channel in a common polarization-based QKD source that employs a Faraday mirror for phase stabilization. We apply the recently developed numerical proof technique from [Phys. Rev. A \textbf{99}, 062332 (2019)] to quantify the sensitivity of the secret key rate to the quantum optical model for the side-channel, and to develop strategies to mitigate the information leakage. In particular, we find that the MDI three-state and BB84 protocols, while yielding the same key rate under ideal conditions, have diverging results in the presence of a side-channel, with BB84 proving more advantageous. While we consider only a representative case example, we expect the strategies developed and key rate analysis method to be broadly applicable to other leaky sources.
\end{abstract}

\maketitle

\section{Introduction}\label{sec:intro}

Quantum key distribution (QKD) allows two separated parties to have information-theoretic secure communication by leveraging the principles of quantum mechanics. In practice,
QKD systems suffer from imperfections and open up side-channel
attacks. In particular, detectors are the weakest link in QKD.
Luckily, measurement-device-independent (MDI) QKD removes all side-channels in detectors \cite{Lo2012}. Nonetheless, imperfections of the quantum state source continue to threaten the security of MDI protocols. To partly address this challenge, the loss tolerant protocol \cite{loss_tol} provides a proof technique for dealing with state preparation flaws, with extensions of the proof available to account for the decoy state method \cite{LT_exp}, and mixed states \cite{bourassa2020loss}. The method from \cite{PhysRevA.88.062322} can also treat flawed sources, under the condition the encoded signals remain confined to a qubit space. Unfortunately, these methods to deal with state preparation flaws only account for systematic errors in the two-dimensional (qubit) degree of freedom that Alice and Bob intentionally encode, meaning these techniques are not sufficient to account for source side-channels. In an effort to generalize to sources that do not output idealized qubits, recent security proof techniques have been developed to deal with sources leaking decoy state parameters \cite{tamaki2016decoy,wang2021measurement}, and encoding information, with analytic approaches given in \cite{pereira2020quantum,Navarrete2021} and numerical techniques in \cite{Pereira2019,Primaatmaja2019}.

With such security proof techniques now available, it is time they be applied to develop practical strategies for MDI QKD protocols employing realistic sources, bringing closer together the gap between idealized security proofs and experimental realities. In this paper, we study a common optical source for polarization-based MDI QKD which relies on a Faraday mirror for polarization stabilization \cite{lucio2009proof,tang2014experimental,LT_exp,wang2016experimental,li2018secure,moschandreou2021experimental, Comandar2016}. We determine that this experimental setup introduces a passive side-channel(i.e. not introduced by Eve) due to leakage light between optical pulses being unintentionally modulated in a time-dependent manner, a loophole that has not been identified in the literature to the best of our knowledge. While some work has been done on computing secure key rates in the presence of passive side-channels---the authors of \cite{duplinskiy2021bounding} establish a lower bound on the key rate in the presence of passive side-channels using signal indistinguishability---most previous works focus on active side channels (side channels introduced by Eve), such as Trojan horse attacks \cite{Gisin2006,Lucamarini2015}, despite passive side channels being harder to avoid. Note that the passive/active dichotomy here refers to whether Eve introduces the side channel, as opposed to the passive/active distinction made for optical elements \footnote{In fact, the passive side-channel we study in this paper is introduced by active optical elements controlled by Alice and Bob.}. For an exhaustive review of hacking strategies for various QKD systems, including via side-channels, see Table I of \cite{XuRMP2020}.

In this particular passive side channel case, we are faced with the seemingly daunting task of incorporating optical states distributed over a continuum of temporal modes into a security proof. However, we find that the versatile proof technique from \cite{Primaatmaja2019} can be employed even in this scenario, a modest extension of its already wide applicability. As a numerical approach, the technique from \cite{Primaatmaja2019} is particularly well-suited to our task, as it allows one to integrate detailed information about the initial states sent by Alice and Bob (including the time-dependent side-channels) and all the observed detection statistics in the protocol as constraints in the security proof.

Using the Faraday mirror source as a representative case example of sources with time-varying side-channels, we calculate the secret key rate under various assumptions and scenarios to better determine strategies for mitigating the information leaked via the side-channel. In particular, we investigate how the model for the state of the side-channel can have a significant impact on the amount of key generated, reinforcing the importance of carefully characterizing the optical output of the source. We present a few practical strategies for increasing the key rate, such as using all available detection statistics, sending more states than what would be required in the ideal protocol, and optimizing the choice of which test state to send from the Bloch sphere. As part of this analysis, we determine that while the MDI three-state protocol \cite{fung2006security,loss_tol} yields the same key rates as the MDI BB84 protocol \cite{bennett1984proceedings,Lo2012} in the ideal case of no side-channel, in the presence of leakage light these two protocols diverge, with BB84 being the more advantageous choice. 

We briefly summarize why we will employ the numerical approach of \cite{Primaatmaja2019} based on semi-definite programming to address the side-channel problem. Note that some other approaches such as the loss-tolerant protocol approach \cite{loss_tol} and uncharacterized qubit approach \cite{PhysRevA.88.062322} cannot be applied to the side-channel problem because those approaches assume the optical source sends out a qubit and such a qubit assumption is violated by side channels.

On the other hand, approaches such as \cite{Pereira2019} and the reference state approach \cite{pereira2020quantum,Navarrete2021} do work for side channels and non-qubit sources. Nonetheless, we find that  technique from \cite{Pereira2019} relaxes the task of bounding the phase error to a linear program and, therefore, it gives a less strict result than using the approach in \cite{Primaatmaja2019}. As for the reference state technique \cite{pereira2020quantum,Navarrete2021}, we find that it gives a worse key rate for the MDI version of BB84 protocol in the presence of side-channels than the approach in \cite{Primaatmaja2019}. For these reasons, we find that the approach in \cite{Primaatmaja2019} is highly suitable for addressing the side channel problem.

The structure of this paper is as follows: in Section \ref{sec:background}, we review the security proof technique from \cite{Primaatmaja2019} and compare it to competing proof techniques \cite{loss_tol,PhysRevA.88.062322,Pereira2019,pereira2020quantum,Navarrete2021} to justify our choice of approach. Then, in Section \ref{sec:case_ex}, we study the case example of a polarization-based MDI QKD setup that employs a Faraday mirror in the transmitter for polarization stabilization; here, we identify a side-channel arising from the time-varying polarization modulation of the leakage light between optical pulses. This polarization modulation is correlated with the polarization encoding of Alice and Bob's signals. We determine that the proof technique from \cite{Primaatmaja2019} can be applied to treat practical sources with time-varying optical signals.

Finally, in Section \ref{sec:results}, we provide key rate results for various protocol scenarios. We determine the impact of the source model, finding that accounting for the time-dependent nature of the side-channel provides a benefit to the key rate over more pessimistic, rudimentary models, thus demonstrating the importance of careful side-channel characterization. Additionally, we find a divergence between the three-state and BB84 protocols in the presence of a side-channel, and determine that the extra state sent in BB84, while redundant under ideal conditions, is able to better mitigate the source information leakage.

\section{Background}\label{sec:background}

To understand the dependence of the secret key rate on the side-channel, we first provide some background on the components that are required for the key rate calculation. In Section \ref{subsec:proof} we review the proof technique from \cite{Primaatmaja2019}, and
in Section \ref{subsec:pf_comp}, we compare our choice of proof technique to other potential options we could have chosen. As our key rate calculations rely on the decoy state protocol, we provide a review of its use in MDI QKD in Appendix \ref{app:decoy_lin_prog}.

\subsection{Security Proof Technique Based on Semidefinite Programming}\label{subsec:proof}

Semidefinite programs (SDPs) are a class of convex optimization problems that can be written in the form:
\begingroup
\setlength{\tabcolsep}{10pt} 
\renewcommand{\arraystretch}{1.5} 
\begin{center}
\begin{tabular}{ l l }
\texttt{maximize} & $f_0(G) = \text{Tr}(A_0 G)$  \\ 
\texttt{s.t.}  & $f_i (G) = \text{Tr}(A_i G)\geq b_i,\ i=1,\dots,m$\\
 & $G\succeq 0$
\end{tabular}
\end{center}
\endgroup
\noindent where $G$ is a positive semidefinite (PSD) matrix (i.e. has non-negative eigenvalues) whose elements form the optimization variables of the problem. $f_0:\mathbb{R}^n\rightarrow\mathbb{R}$ is the objective function we seek to maximize. $f_i:\mathbb{R}^n\rightarrow\mathbb{R}$ are the constraint functions, and $b_i$ are the constraint bounds. Importantly, the objective and the constraint functions are all linear functions of the elements of $G$, with the coefficients of the linear functions contained in the matrices $A_i$. SDPs are increasingly being used for QKD security proofs \cite{Coles2016, Winick2018reliablenumerical,  Islam2018, Wang2019, Islam2019, Primaatmaja2019,tan2019computing,zhang2020securing,george2021numerical}, due in part to the availability of fast and mature numerical implementations of solvers \cite{cvxpy,cvxpy_rewriting}.

In \cite{Primaatmaja2019}, the authors present a versatile numerical proof technique based on semidefinite programs (SDPs) for MDI QKD protocols. The objective function of the SDP is the phase error of the key rate formula---either Shor-Preskill \cite{Shor2001} or the GLLP version for decoy states \cite{GLLP2003} (see Eq. \eqref{eq:KR_formula})---meaning an optimal solution provides a secure lower bound. The constraints are provided by the detection statistics of the protocol, as well as initial state information, which is especially useful since it allows more experimental information to be used to quantify security. Here, we review the proof technique from \cite{Primaatmaja2019} so that later we can apply it to the case of an MDI QKD protocol with a source side-channel.

To begin, we distinguish between the full optical states that Alice and Bob send to Charlie from the components of those states from which we will derive security. We will refer to the components of the optical states used to derive security as the \textit{signal} states. As an example, in an ideal decoy state protocol, the full optical states are phase-randomized WCPs with different intensities, while the signal states used to derive security are the single-photon components. We review the decoy state protocol in Appendix \ref{app:decoy_lin_prog}. Alternatively, in an ideal phase-encoding protocol that does not use decoy states, the optical state and the signal state are one and the same.

Let the signal states that Alice prepares be denoted by $\ket{\psi^{i}_{x}}_A$, where $(i,x)$ indicate her choice of basis and bit value. Analogously, we can write Bob's states as $\ket{\varphi^{j}_{y}}_B$, with his basis and bit choice given by $(j,y)$. Like in \cite{Primaatmaja2019}, we will assume that the signal states are pure states; however, a path to treating mixed states is available via the technique from \cite{bourassa2020loss}. Note that, even though phase-randomized WCPs in an ideal decoy state protocol are mixed states, the signal states (single photon components) are pure states. At a high level, in an MDI QKD protocol, Alice and Bob send their signal states to Eve, who in turn makes an announcement $z$ conditioned on a measurement she may or may not execute faithfully. For the sake of simplicity, we assume $z=P,F$, corresponding to a binary \texttt{pass} or \texttt{fail} outcome, but this can be generalized to account for more announcements. 

Since quantum mechanics obeys unitary evolution, we can write the evolution of the joint state as:
\begin{equation}\label{eq:evol_to_Eve}
    \ket{\psi^{i}_{x}\varphi^{j}_{y}}_{AB} \rightarrow \sum_{z=P,F}\ket{e^{i,j}_{x,y,z}}_E\ket{z}_Z
\end{equation}
where $\ket{e^{i,j}_{x,y,z}}_E$ are sub-normalized vectors that can be used to completely characterize Eve's state. Were we to know the states $\ket{e^{i,j}_{x,y,z}}_E$, we would have full knowledge of Eve's information about the key, and in turn could calculate the key rate exactly. Alas, the exact state of Eve's system is generally unknown; however, we can impose constraints on the state vectors $\ket{e^{i,j}_{x,y,z}}_E$. As we will review now, one can frame the secret key rate calculation as a semidefinite program, where the PSD matrix used as an optimization variable is the Gram matrix of Eve's vectors which we denote $G_E$. We recall that the elements of a Gram matrix for a set of vectors are all the pairwise inner products of the vectors, meaning $G_E$ has elements $\braket{e^{i',j'}_{x',y',z'}|e^{i,j}_{x,y,z}}_E$. Gram matrices are always PSD. 

The first type of constraint comes from the unitary evolution of states in quantum mechanics; namely, the inner product structure of the initial states must be preserved in the final states \cite{Primaatmaja2019}. If Alice and Bob each have $n_A$ and $n_B$ basis choice settings, each basis choice associated with two bit choices, then the inner product constraint yields $(n_A \times n_B\times2\times2)^2$ constraints of the form:
\begin{equation}\label{eq:inner_prod}
        \braket{\psi^{i'}_{x'}\varphi^{j'}_{y'}|\psi^{i}_{x}\varphi^{j}_{y}}_{AB} = \sum_{z}\braket{e^{i',j'}_{x',y',z}|e^{i,j}_{x,y,z}}_E,
\end{equation}
where the fact that the announcements are classical means $\braket{z|z'}_Z=\delta_{z,z'}$. Note that these constraints are linear in the elements of $G_E$

The next type of constraint on $G_E$ comes from the observed detection statistics \cite{Primaatmaja2019}. Let the probability of Eve announcing a successful detection event, conditioned on Alice and Bob having chosen basis and bit choices $(i,j,x,y)$ be denoted by $p_{\texttt{pass}}^{i,j,x,y}$. In an ideal decoy state protocol, $p_{\texttt{pass}}^{i,j,x,y}\equiv p_{\texttt{pass},1,1}^{i,j,x,y}$, since the single photon components are the signal states. This can be related to the elements of $G_E$ as follows:
\begin{equation}\label{eq:det_prob}
    p_{\texttt{pass}}^{i,j,x,y} = \braket{e^{i,j}_{x,y,P}|e^{i,j}_{x,y,P}}_E
\end{equation}
If the signal states are the same as the full optical states, then $p_{\texttt{pass}}^{i,j,x,y}$ would be directly observable in practice. Alternatively, if one is performing a decoy state MDI QKD protocol, then, as reviewed in Appendix \ref{app:decoy_lin_prog}, one first uses the statistics of the full optical states in a linear program to establish upper and lower bounds on $p_{\texttt{pass}}^{i,j,x,y}$:
\begin{equation}\label{eq:ineq_det_cons}
    p_{\texttt{pass},L}^{i,j,x,y} \leq p_{\texttt{pass}}^{i,j,x,y} \leq p_{\texttt{pass},U}^{i,j,x,y}.
\end{equation}
Thus, depending on the protocol, one either obtains $4n_A n_B$ equality constraints, or $8 n_A n_B$ inequality constraints on $G_E$. Like the previous set of constraints, these are also linear in elements of $G_E$.

So far, we have identified the Gram for Eve's system $G_E$ as a PSD matrix, as well as various linear constraints on its elements. We now review how to write the phase error as the objective function of an SDP. To start, we will assume that the basis choice $(i,j)=(0,0)$ corresponds to the key generation basis. Moving to a virtual picture, we can think of Alice and Bob's signal states being entangled with virtual qubits $\bar{A}$ and $\bar{B}$ that they keep in their lab:
\begin{equation}
    \ket{\Psi_{virt}}_{\bar{A}\bar{B}AB} = \sum_{x,y=0}^{1} \ket{x}_{\bar{A}}\ket{y}_{\bar{B}}\ket{\psi^{0}_{x}\varphi^{0}_{y}}_{AB},
\end{equation}
where measurement of $\bar{A}\bar{B}$ in the computational basis yields the bit values of the secret key. Let the virtual state evolve to $\ket{\Psi_{virt}}_{\bar{A}\bar{B}AB} \rightarrow \ket{\Gamma}_{\bar{A}\bar{B}EZ}$ with:
\begin{equation}
    \ket{\Gamma}_{\bar{A}\bar{B}EZ} = \sum_{x,y=0}^{1} \ket{x,y}_{\bar{A}\bar{B}}\sum_{z=P,F}\ket{e^{0,0}_{x,y,z}}_E\ket{z}_Z,
\end{equation}
since we used Eq. \eqref{eq:evol_to_Eve}. 

Through the process of sending their key generation signal states to Eve (who conducts a measurement), as well as postselecting on $z=P$, Alice and Bob end up with a mixture of Bell states in the $\bar{A}\bar{B}$ virtual qubits. The virtual picture therefore allows us to formally define the phase error rate that characterizes security in the key rate in Eq. \eqref{eq:KR_formula}. Assuming, without loss of generality, that the target Bell state of the protocol is $\ket{\Phi^+}_{\bar{A}\bar{B}} = \frac{1}{\sqrt{2}}(\ket{00}_{\bar{A}\bar{B}}+\ket{11}_{\bar{A}\bar{B}})$, then the phase error rate is defined to be the probability that the $\bar{A}\bar{B}$ virtual qubits held by Alice and Bob end up in Bell states with the incorrect phase:
\begin{equation}
\begin{split}
    e_{ph} &= \frac{\bra{\Gamma} \left(\mathbb{M}^{-}_{\bar{A}\bar{B}}\otimes \ket{P}\bra{P}_Z\right)\ket{\Gamma}_{\bar{A}\bar{B}EZ}}{\bra{\Gamma} (\ket{P}\bra{P}_Z)\ket{\Gamma}_{\bar{A}\bar{B}EZ}}\\
    &= \frac{1}{2} - \frac{Re\left(\braket{e^{0,0}_{0,0,P}|e^{0,0}_{1,1,P}}_E+\braket{e^{0,0}_{0,1,P}|e^{0,0}_{1,0,P}}_E\right)}{\sum_{x,y}p_{\texttt{pass}}^{0,0,x,y}},
\end{split}
\end{equation}
with
\begin{equation}
    \mathbb{M}^{-}_{\bar{A}\bar{B}} = (\ket{\Phi^-}\bra{\Phi^-}+\ket{\Psi^-}\bra{\Psi^-})_{\bar{A}\bar{B}}.
\end{equation}
Note that $e_{ph}$ is a linear function of the elements of $G_E$ as required for an SDP. With the additional constraint that $0 \leq e_{ph} \leq \nicefrac{1}{2}$, such that we are within the region where the binary entropy function increases monotonically, then we can maximize $e_{ph}$ via an SDP and determine a secure lower bound on the key rate using the Shor-Preskill formula \cite{Shor2001} or the key rate from \eqref{eq:KR_formula}. Note that in a decoy state protocol without leakage light, $e_{ph}=e_{ph,1,1}$, since the signal states correspond to the single photons components.

In summary, \cite{Primaatmaja2019} provides a method for obtaining a secure lower bound on the key rate using an SDP of the form:
\begingroup
\setlength{\tabcolsep}{4pt} 
\renewcommand{\arraystretch}{2} 
\begin{center}
\begin{tabular}{ l l }
\texttt{maximize} & $e_{ph}$  \\ 
\texttt{s.t.}  & $\braket{\psi^{i'}_{x'}\varphi^{j'}_{y'}|\psi^{i}_{x}\varphi^{j}_{y}}_{AB} = \sum_{z}\braket{e^{i',j'}_{x',y',z}|e^{i,j}_{x,y,z}}_E$\\
 &     $p_{\texttt{pass},L}^{i,j,x,y} \leq \braket{e^{i,j}_{x,y,P}|e^{i,j}_{x,y,P}}_E \leq p_{\texttt{pass},U}^{i,j,x,y}$\\
 & $0 \leq e_{ph} \leq \nicefrac{1}{2}$\\
 & $G_E\succeq 0$.
\end{tabular}
\end{center}
\endgroup
\noindent With this technique in hand, we apply it to the case example of an MDI QKD source with a side-channel, which we will describe in the next section. First, however, we justify our choice of proof technique by comparing it with competing methods.

\subsection{Comparison to competing proof techniques}\label{subsec:pf_comp}

Given our review of the numerical approach from \cite{Primaatmaja2019}, it is worth comparing with competing proof techniques for MDI QKD to see why the approach we have chosen is well-suited to the problem we wish to study. The loss tolerant protocol \cite{loss_tol} and the proof technique from \cite{PhysRevA.88.062322} are both leading techniques for quantifying security in the presence of state-preparation flaws, the former requiring knowledge of the initial states, while the latter can simply use the detection statistics. However, in both techniques, one needs to assume that the optical source is outputting a qubit state, meaning they are insufficient to treat scenarios involving source side-channels, since the extra state sent with the encoded qubit, e.g. an optical coherent state, can easily break the assumption that the source only outputs states from a two-dimensional Hilbert space.

The numerical proof technique developed in \cite{Coles2016,Winick2018reliablenumerical} also uses SDPs to compute the secret key rate; however, they work directly with the Devetak-Winter key rate formula \cite{Devetak2005}, as opposed to the Shor-Preskill key rate \cite{Shor2001}. Working with the Devetak-Winter key rate requires a series of relaxations, and solving a series of SDPs, which is more cumbersome and numerically slower than a direct calculation of the phase error.

Two generalizations of the loss tolerant protocol have been developed to deal with non-qubit sources, in part for the purpose of studying source side-channels: the technique from \cite{Pereira2019} and the reference state technique \cite{pereira2020quantum, Navarrete2021}. These techniques are directly comparable to \cite{Primaatmaja2019} as they all use the Shor-Preskill key rate, with the core task of the proof being to find an upper bound on the phase error rate. While shown to perform more poorly than the reference state technique \cite{pereira2020quantum}, the technique from \cite{Pereira2019} may be the most directly comparable to \cite{Primaatmaja2019}. Both methods allow one to consider an arbitrary number of initial states that are not confined to a qubit space, and involve using all observed detection statistics. As we show in Appendix \ref{app:pereira_comp}, however, the approach from \cite{Pereira2019} relaxes the task of bounding the phase error to a linear program. The constraints of the linear program are provided by the detection probabilities and the initial states; however, fewer constraints are provided by the initial states than the approach from \cite{Primaatmaja2019}, since the overlaps between states with different $(i,j,x,y)$ are not considered. Moreover, linear programs are a class of convex optimization problems contained within SDPs, meaning the constraints on the optimization variables used to compute the phase error are less strict than solving the full SDP as done in \cite{Primaatmaja2019}. Thus, we expect the key rates provided by the SDP numerical approach to be greater or equal to those calculated using the technique from \cite{Pereira2019} in general. In Appendix \ref{app:pereira_comp}, we provide the illustrative example of the three state protocol with a side channel to explicitly demonstrate the key rates provided by the SDP method outperform those from \cite{Pereira2019}.

Finally, while the reference state technique is a purely analytic approach, a disadvantage is that it can currently only treat the case when Alice and Bob each send three states. The crux of the technique is to consider hypothetical detection statistics and phase error stemming from a fictitious set of reference states, and then bound the actual phase error of the protocol using the real detection statistics and the deviation between the reference and real states \cite{pereira2020quantum,Navarrete2021}. In particular, in \cite{pereira2020quantum}, the strategy for treating protocols involving four states, such as BB84, is to consider random alternation between two three-state protocols, where each of the X-basis BB84 states act as the third state. While the three-state protocol and BB84 yield the same key rates when the initial states are qubits \cite{loss_tol}, one of the observations that this work will provide is that information from the seemingly redundant fourth state of BB84 can provide extra constraints to boost the key rate in the presence of a source side-channel (i.e. when the qubit assumption is broken). Thus, a downside of using the reference state technique for protocols involving more than three states is that the key rate calculation will only ever be constrained by the statistics and initial states of three out of the four states, which leaves valuable information on the table, at the cost of higher key rate.

\section{Source Side-Channels: A Case Example}\label{sec:case_ex}

Having reviewed the necessary components for the security proof, we now study a case example of an MDI QKD source with a side-channel. In Section \ref{subsec:side-chan-origin}, we identify a novel, time-dependent passive source side-channel which occurs when using a Faraday mirror for stable electro-optic bit modulation \cite{lucio2009proof,tang2014experimental,LT_exp,wang2016experimental,li2018secure,moschandreou2021experimental, Comandar2016}. We provide quantum optical modelling of the side-channel in Section \ref{subsec:QO_model}, and in Section \ref{subsec:apply-pf}, we link the model to the security proof technique described in Section \ref{sec:background} to assess its impact on security while taking its time-dependent nature into account.

\subsection{Origin of the Side-Channel}\label{subsec:side-chan-origin}

\begin{figure}
    \centering
    \includegraphics[width=\columnwidth]{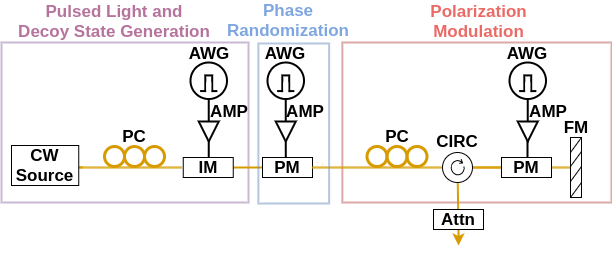}
    \caption{Experimental setup for polarization encoding MDI-QKD transmitter. An intensity modulator is used to create pulsed light (including decoy states) from a continuous wave light source. Then, the pulses go through a phase randomization unit, followed by a polarization modulation/encoding unit.  PC -  polarization controller, IM - intensity modulator, AMP - voltage amplifier, AWG - arbitrary waveform generator, PM - phase modulator, CIRC - optical circulator, FM - Faraday mirror, Attn - optical attenuator.}
    \label{fig:exp_setup}
\end{figure}

Several polarization and phase encoding implementations of MDI, prepare-and-measure, and plug-and-play QKD make use of an electro-optic phase modulator and Faraday mirror for optical bit modulation \cite{tang2014experimental,Zhou2003}. The Faraday mirror is added, as shown in Figure \ref{fig:exp_setup}, to remove the temperature dependence of the phase modulator \cite{lucio2009proof,tang2014experimental,LT_exp,wang2016experimental,li2018secure,moschandreou2021experimental}. Optical pulses first travel forward through the PM, co-propagating with a voltage pulse. During this first trip, they experience both voltage and unintentional temperature induced phase modulation. After reflection from the Faraday mirror, the pulses travel back through the PM, such that they do not collide with any counter-propagating voltage pulses. Hence, during this second trip, they only experience temperature induced phase modulation. Although the usage of a Faraday mirror drastically reduces state preparation flaws, we found that it creates a source side-channel. 

This side-channel occurs due to the presence of weak light leakage between the optical pulses into which bits are encoded. These optical pulses are carved out from continuous wave light using an electro-optic intensity modulator (IM). Due to the finite extinction ratio of pulses that can be created with an IM, the presence of weak light leakage is inevitable. Of course, the phase of this leakage light is not intentionally modulated. In other words, no voltage is applied to the phase modulator as this light travels through it for the first time. However, after reflection from the Faraday mirror, some of this leakage light would inevitably collide with a counter-propagating voltage pulse within the phase modulator. Therefore, this leakage light would experience unintentional voltage-induced phase modulation, forming a source side-channel whose impact on security must be quantified. 

\begin{figure}
    \centering
    \includegraphics[width=\columnwidth]{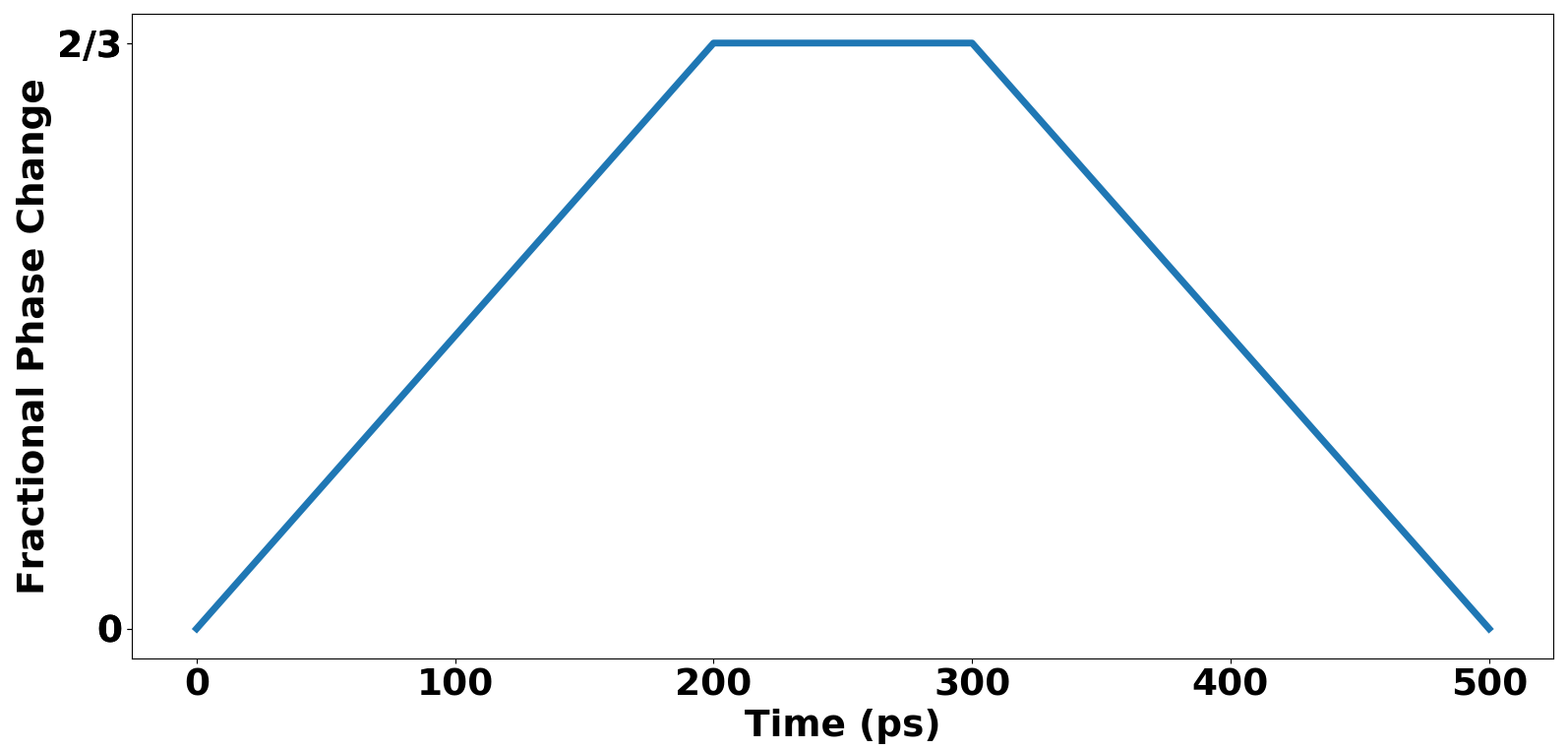}
    \caption{Time dependent fractional phase change applied to the leakage light. Fraction is with respect to the phase change applied to the corresponding encoded light.}
    \label{fig:leakage_pol}
\end{figure}

We find the modulation of the leakage light to be time-dependent, as it travels in the opposite direction of the voltage pulse. It is also dependent on the voltage pulse shape used for phase modulation and the phase modulator electrode length. For our particular setup, the time dependence is shown in Figure \ref{fig:leakage_pol}. Refer to Appendix \ref{app:fig_derivation} for further details on how Figure \ref{fig:leakage_pol} was derived.

Several simplifying assumptions regarding the optical and leakage signals are made when performing the security analysis. 
\begin{enumerate}
    \item In our particular setup (see Figure \ref{fig:exp_setup}), polarization encoding occurs after setting the decoy state intensity and performing phase randomization. Hence, we assume that the decoy state intensity and phase randomization of optical pulses (encoded signals) are independent of their polarization encoding.
    \item We assume that the voltage pulses delivered to the polarization modulation PM are square pulses, such that the polarization of encoded pulses are time-independent. 
    \item In our experimental setup, there is a minimal correlation between the decoy state intensity and leakage light intensity. This correlation stems from the tails of the pulse shaping voltage pulses, which are small compared to the full duration of leakage light between pulses. Therefore, we assume that leakage signals carry no information about the decoy state intensity setting of encoded signals.
    \item We assume that the voltage pulses strictly overlap with the optical pulses within the phase randomizing PM. Hence, the leakage signals carry no information about the phase randomization of encoded signals.
    \item We assume that the leakage intensity is uniform in time, but the method could also easily treat time-varying intensity. 
\end{enumerate}

Note that these assumptions could be broken and could be incorporated into the security analysis using techniques from \cite{wang2021measurement}. However, in our particular experimental setup, they are not the leading-order source of information leakage, which we take to be the unintentional polarization modulation of the leakage light after reflection in the Faraday mirror. We can now proceed to model the quantum state of the source's output light.

\subsection{Quantum Optical Modelling}\label{subsec:QO_model}
To proceed with the security proof technique from Section \ref{subsec:proof}, we must first model the states transmitted by the source so that we can compute the inner products of the signal states. The full optical state can be written as a separable state of Alice’s and Bob’s transmitted states: $\rho_A^{k,i,x} \otimes \sigma_B^{l,j,y}$. The parameters $(k,l)$ refer to their decoy state intensity settings, $(i, j)$ refer to their basis choice, and $(x,y)$ refer to their bit choices. The basis and bit choices impact the polarization set with the polarization phase modulator, and are independent from the choice of intensity setting.

Alice's state can be further broken down into the side-channel state which represents the leakage light, and the encoding state which represents the optical pulses into which the basis and bit information are intentionally encoded:
\begin{equation}
    \rho_A^{k,i,x} = \rho_{enc}^{k,i,x} \otimes \rho_{leak}^{i,x}.
\end{equation}
We assume that the intensity modulation and phase randomization are timed with the optical pulses, and that in between pulses when the leakage light is passing through these modules, no phase randomization is applied, and the intensity modulation attenuates the light as much as possible to constant minimum but non-zero intensity. This has several consequences: first, this means that only $\rho_{enc}^{k,i,x}$ carries information about the intensity setting, and that $\rho_{leak}^{i,x}$ carries information about neither the random phase nor the intensity. This means we can treat $\rho_{enc}^{k,i,x}$ as a perfectly phase-randomized WCP just as in an ideal MDI decoy state protocol without leakage light. Second, our assumptions mean that we can treat $\rho_{leak}^{i,x}$ as a pure state:
\begin{equation}
    \rho_{leak}^{i,x} = \ket{\chi^{i}_{x}}\bra{\chi^{i}_{x}}_{leak}.
\end{equation}

We now move to model the polarization module of the source and the time-dependent nature of the side-channel state. When $\rho_{enc}^{k,i,x}$ passes through the phase modulator, the controlling voltage pulse is timed with the optical signal such that the resulting polarization, determined by settings $(i,x)$, is time-independent across the length of the optical pulse. By contrast, because $\rho_{leak}^{i,x}$ is travelling in the opposite direction, it acquires a time-dependent polarization. Let the creation operator for a photon at time $t$ with polarization angles $\theta^i_x(t)$ and $\phi^i_x(t)$ be given by:
\begin{equation}
    a^\dagger_{t,i,x} = \cos[\theta^i_x(t)] a_{t,H}^\dagger + \sin[\theta^i_x(t)] e^{i\phi^i_x(t)} a_{t,V}^\dagger,
\end{equation}
where $H$ and $V$ denote the horizontal and vertical polarization modes, with the raising and lowering operators satisfying $[a_{t,m},a^\dagger_{t',m'}]=\delta(t-t')\delta_{m,m'}$, $m=H,V$. As non-phase-randomized laser light, the leakage light at a given instant in time $t$ can be treated as a coherent state with amplitude $\alpha^i_{x}(t)$, meaning over multiple times, the state can be written in general as:
\begin{equation}\label{eq:timedep_coh}
    \ket{\alpha^i_x(t)} = \exp\left\{\int dt[ \alpha^i_x(t) a^\dagger_{t,i,x} - {\alpha^i_x}^*(t) a_{t,i,x}]\right\}\ket{vac}.
\end{equation}
In the source we are studying, we assume the leakage light has an effectively constant intensity $|\alpha_0|^2$ and polar angle $\theta$, while the azimuthal angle $\phi$ changes with time based on the interaction with the phase modulator, as shown in Fig. \ref{fig:leakage_pol}; however, the technique we will apply could easily be used to study time-dependent intensity and polar angles as this would just modify the integral over time used to calculate the inner product between two side-channel states. The state of the leakage light associated with a given pulse is spread over multiple temporal modes, and is given by
\begin{equation}\label{eq:model3}
    \ket{\chi^{i}_{x}}_{leak} = \exp\left[\int_{-\Delta/2}^{\Delta/2} dt\, \alpha_0 a^\dagger_{t,i,x} - \alpha_0^* a_{t,i,x}\right]\ket{vac},
\end{equation}
where $a^\dagger_{t,i,x}$ here denotes the creation operator for a polarized photon with time-independent polar angle $\theta^{i}_{x}$ and time-dependent azimuthal angle $\phi^{i}_{x}(t)$ as in Fig. \ref{fig:leakage_pol}, where the angles depend on Alice's basis and bit choices $(i,x)$. $\Delta$ is the duration of the leakage light that contains encoding information, which from Fig. \ref{fig:leakage_pol} is 500 ps.

In summary, we have that the output state of Alice's source can be treated as a tensor product of a perfectly phase-randomized WCP in the encoded mode with a time-varying pure coherent state in the side-channel mode. The time-varying polarization of the side-channel state depends on the basis and bit values chosen, but not the intensity setting or random phase used for the decoy state method. We can model Bob's source in the same way, denoting his encoded and leakage states by $\sigma_{enc}^{l,j,y}$ and $\ket{\zeta^{j}_{y}}_{leak}$, respectively. In the next section, we use these assumptions about the source, in connection with the decoy state method and proof technique reviewed in Section \ref{sec:background}, to build up the security proof for this MDI QKD source.

\subsection{Applying the proof technique}\label{subsec:apply-pf}

Given the model of the optical source, we can now connect it with the decoy state method reviewed in Appendix \ref{app:decoy_lin_prog} and security proof technique from Section \ref{subsec:proof}. To start, since the intentionally encoded states $\rho_{enc}^{k,i,x}\otimes\sigma_{enc}^{l,j,y}$ can still be treated as phase-randomized WCPs, and since the side-channel states $\ket{\chi^{i}_{x}}_{leak}\otimes\ket{\zeta^{j}_{y}}_{leak}$ carry no information about the decoy state intensity or random phase, we are able to use the decoy state method with only modifications to how we interpret the detection probabilities obtained by solving the linear programs. 

The photon number distribution of the states $\rho_{enc}^{k,i,x}\otimes\sigma_{enc}^{l,j,y}$ follows the form from Eq. \eqref{eq:decoy_photon_dist}; however, when considering the linear equations provided by the detection probabilities in Eq. \eqref{eq:tot_det_prob}, $p_{\texttt{pass},m,n}^{i,j,x,y}$ now refers to the probability a round passes given that Alice sent the $m$-photon component of the state $\rho_{enc}^{k,i,x}$, that Bob sent the $n$-photon component of the state $\sigma_{enc}^{l,j,y}$, \textit{and} that they together sent the leakage states $\ket{\chi^{i}_{x}}_{leak}\otimes\ket{\zeta^{j}_{y}}_{leak}$. Note that the state of the leakage light does not depend on the number of photons Alice and Bob sent, just on the polarization encoding choice, so we can still use $m,n$ to label the variable, even though it does not strictly refer to Fock states anymore. Moreover, since the leakage states are independent of the intensity choice settings, $p_{\texttt{pass},m,n}^{i,j,x,y}$ remains independent of $(k,l)$. Solving the linear program in Appendix \ref{app:decoy_lin_prog}, Alice and Bob retrieve, for each basis and bit setting $(i,j,x,y)$, bounds on the probabilities $p_{\texttt{pass},1,1}^{i,j,x,y}$ that Eve will announce a successful detection event given that they each sent a single photon in the encoded mode \textit{along with} the associated side-channel state.

Interpreting $p_{\texttt{pass},1,1}^{i,j,x,y}$ as coming from both the single photon components of the encoded mode \textit{and} from the leakage light, means that the definition of the signal states in this protocol no longer refers just to the single photon components of the encoded mode, as is the case for an ideal decoy state protocol. Connecting to Eq. \eqref{eq:evol_to_Eve}, Alice and Bob's signal states are given by:
\begin{equation}
    \ket{\psi^{i}_{x}\varphi^{j}_{x}}_{AB} = \ket{\psi^{i}_{x}\varphi^{j}_{y}}_{enc}\otimes\ket{\chi^{i}_{x}\zeta^{j}_{y}}_{leak},
\end{equation}
where $\ket{\psi^{i}_{x}\varphi^{j}_{x}}_{enc}$ refers to the single photon components of the phase-randomized WCPs $\rho_{enc}^{k,i,x}\otimes\sigma_{enc}^{l,j,y}$. With these as the signal states, the detection probability constraints from Eq. \eqref{eq:det_prob} employ the probabilities $p_{\texttt{pass},1,1}^{i,j,x,y}$ which come from the signal states, i.e. the leakage light and the single photon component of the encoded mode. Additionally, the inner product constraints from Eq. \eqref{eq:inner_prod} now include the inner products of the states of the leakage light; this reaffirms the versatility of the proof technique we are employing, since the constraints coming from the continuous-variable, time-dependent leakage light states can be coarse-grained down to their inner products, which form a finite-dimensional Gram matrix. The optimization variables (the elements of Eve's Gram $G_E$) and the objective function (the phase error rate) do not change; however, since the constraints will be affected by the presence of leakage light, the resulting key rate will certainly change.

With a model for the source, and how it connects to the decoy state method and the proof technique, we can now move to calculate the secret key rate for various scenarios and protocols.

\section{Key Rate Results}\label{sec:results}

Having reviewed the main components required for the security proof technique in Section \ref{sec:background}, and having studied a case example of an MDI QKD source with a side-channel in Section \ref{sec:case_ex}, we now calculate key rates under different conditions. In Section \ref{subsec:sim_details}, we provide the details of how our simulations were performed. Then, in Section \ref{subsec:model_comp}, we see how the key rate can change depending on the model chosen for the side-channel, including the time-dependent state we derived in the previous section, highlighting the need for careful side-channel characterization. In Section \ref{subsec:3vs4} we explore a strategy for extracting higher key rates by sending states that would be redundant under ideal conditions but which help in the presence of a side-channel, concluding that the three-state protocol and BB84, which yield equivalent key rates in the ideal case of no leakage light, have different key rates when a side-channel is present. In Appendix \ref{app:more_KR} we examine two more strategies for boosting the key rate by using all mismatch statistics and by modifying the polarization angle of the states sent. These simulations should be used to better inform the choice of protocol when working with realistic sources like the one we are studying. While having only considered a specific source with a non-trivial side-channel, we expect the conclusions drawn to be broadly applicable to any leaky source; namely, we emphasize the importance of side-channel characterization, and determine which protocol parameters (e.g. number of states sent) can lead to key rate improvement.

\subsection{Simulation details}\label{subsec:sim_details}

Recall we have two types of constraints to calculate the key rate: the inner product of the initial states, and the detection probabilities. Before detailing how we simulate these, we first comment on how different aspects of the source model affect these constraints. 

To start, we observe that the single photon component of the encoded mode is the only quantity that affects both the inner product and detection probability constraints. However, whether that single photon state is directly sent from a single photon source, or is a component of a phase-randomized WCP is irrelevant to the inner product constraint; the type of source is only relevant to the detection statistics constraint, since single photons undergoing a lossy channel will provide a different result at a threshold detector than phase-randomized WCPs. 

Next, we assume the usage of gated detectors that would be timed to receive the encoded optical pulses. Hence, in our simulations of the detection events, we assume that the state of the leakage light has no impact on the overall observed detection statistics, since they are in temporal modes that are not picked up by the detectors and the already weak leakage light would be even less bright after the lossy channel. As a consequence, in the calculations we present, the side-channel state only affects the inner-product constraint. We note that the proof technique could easily accommodate the case of detection statistics being affected by the leakage light, as this would just be simulating different values of $Q_{k,l}^{i,j,x,y}$ in Eq. \eqref{eq:tot_det_prob}.

Since different aspects of the source model affect constraints in non-trivial ways, to better understand the key rate resulting from the source described in Section \ref{sec:case_ex}, we provide comparisons to other optical source models. Specifically, we consider:
\begin{itemize}
    \item Single-photon vs. Phase-randomized WCP sources: when calculating the key rate for a given single photon component and side-channel state, i.e. for fixed inner product constraints in Eq. \eqref{eq:inner_prod}, how much is the key rate affected by those signal states being used directly vs. in a decoy state method?
    \item Sensitivity to the side-channel model: we assume in the detection simulations that the side-channel state has no impact on the observed outcomes, so the detection probability constraint in Eq. \eqref{eq:det_prob} remains fixed even if we change the model for the leakage light. In Section \ref{subsec:QO_model}, we provided a model for the source which resulted in a time-dependent coherent state. Were we to change this model, how much does the key rate change?
\end{itemize}
In the sections that follow, we will consider these high-level choices of the model, in addition to varying more practically-rooted parameters like the intensity of the leakage light, plus the number and choice of encoded states sent.

For the choice of side-channel model, we compare three different approaches to treating the state of the leakage light:
\begin{itemize}
    \item Model 1: full encoding information leaked. In this model, we assume the leakage light state is of the form:   
    \begin{equation}
        \ket{\chi_x^i}_{leak} = \sqrt{\epsilon}\ket{vac}+\sqrt{1-\epsilon}\ket{i,x}
    \end{equation}
    with $\braket{i,x|i',x'}=\delta_{i,i'}\delta_{x,x'}$. This model has been used in previous studies of QKD source side-channels \cite{Pereira2019,pereira2020quantum,Navarrete2021}. This model makes a relatively pessimistic assumption, since any non-vacuum component of leakage light provides full-information, while we know, for example, that the single photon component would not be able to unambiguously encode all possible basis and bit choices $(i,x)$.
    \item Model 2: time-independent coherent state. In this model, we assume the leakage light state is of the form:   
    \begin{equation}
        \ket{\chi_x^i}_{leak} = \ket{\beta\cos\theta_x^i}_H\otimes\ket{\beta\sin\theta_x^i e^{i\phi_x^i}}_V.
    \end{equation}
    The angles $(\theta_x^i, \phi_x^i)$ are chosen to coincide with the polarization angles of the encoded mode. This model is more realistic in that we know the leakage light, as laser light, is in a coherent state; however, it does not account for the time-dependent nature of the polarization encoding in the leakage light, which has the opportunity to act to our advantage since not every instant provides Eve with full encoding information. Time-independent coherent state leakage light was considered in the context of Trojan horse attacks in \cite{Primaatmaja2019}.
    \item Model 3: time-dependent coherent state (multiple temporal modes). This model assumes the state from Eq. \eqref{eq:model3}. It is our most accurate model of the source side-channel we introduced in Section \ref{sec:case_ex}. The inner product between two general, time-dependent coherent states is given by:
    \begin{equation}\label{eq:td_inner}
        \braket{\beta(t)|\alpha(t)} = e^{ -\frac{1}{2}\int dt\left[|\beta(t)|^2+|\alpha(t)|^2-2\beta^*(t)\alpha(t)\right]},
    \end{equation}
    which we use to calculate the inner product of the side-channel states in Eq. \eqref{eq:model3}.
\end{itemize}
While we have three different models for the leakage light, we can still associate each of them to a fixed leakage light intensity, $|\alpha|^2$. For Model 1, we can set $\epsilon=e^{-|\alpha|^2}$. In Model 2, we can set $\beta=\alpha$, and in Model 3 we can set $\int dt |\alpha_0|^2 = |\alpha|^2$. This means all the models have the same vacuum probability, i.e. chance of sending no information to Eve, while the non-vacuum components carry varying amounts of information about the basis and bit values.

For the simulation of the detection statistics, in all our simulations we assume detection of a single Bell state using the detector setup from \cite{Lo2012}, with detector efficiency of 50\%, dark count rates of $10^{-6}$ per pulse, and loss in fibre of 0.2 dB/km, with symmetric channel lengths from Alice and Bob to Charlie. To isolate the effect coming from the side-channel, we do not assume any misalignment in the source, but this could easily be added to the detection simulations. When simulating the decoy state method, we have Alice and Bob employ constant intensities of 0.05, 0.1 and 0.6; however, an additional layer of optimization for the decoy state intensities is possible, using our phase error calculation as a subroutine. The detection outcome probabilities $Q^{i,j,x,y}_{k,l}$ for the phase-randomized WCPs were simulated using the method from \cite{Ma2012_mdi_decoy}. All of our calculations are in the asymptotic limit of infinite key length, and in the limit as the sifting rate goes to 1.

\subsection{The Benefits of Side-Channel Characterization}\label{subsec:model_comp}

\begin{figure}
    \centering
    \subfloat[Decoy method]{\includegraphics[width=0.97\columnwidth]{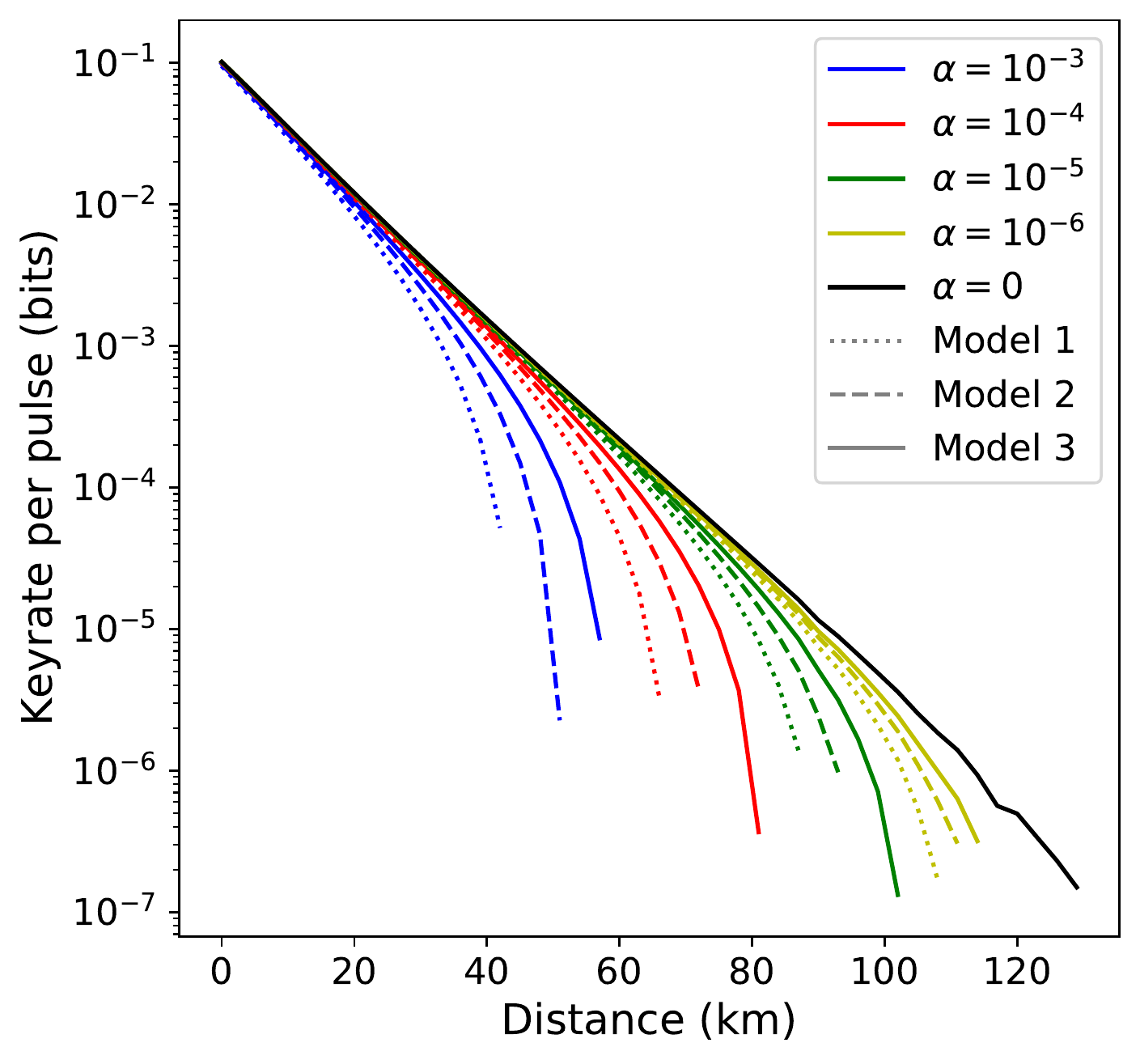}}
    \newline
    \subfloat[Single-photon source]{\includegraphics[width=0.97\columnwidth]{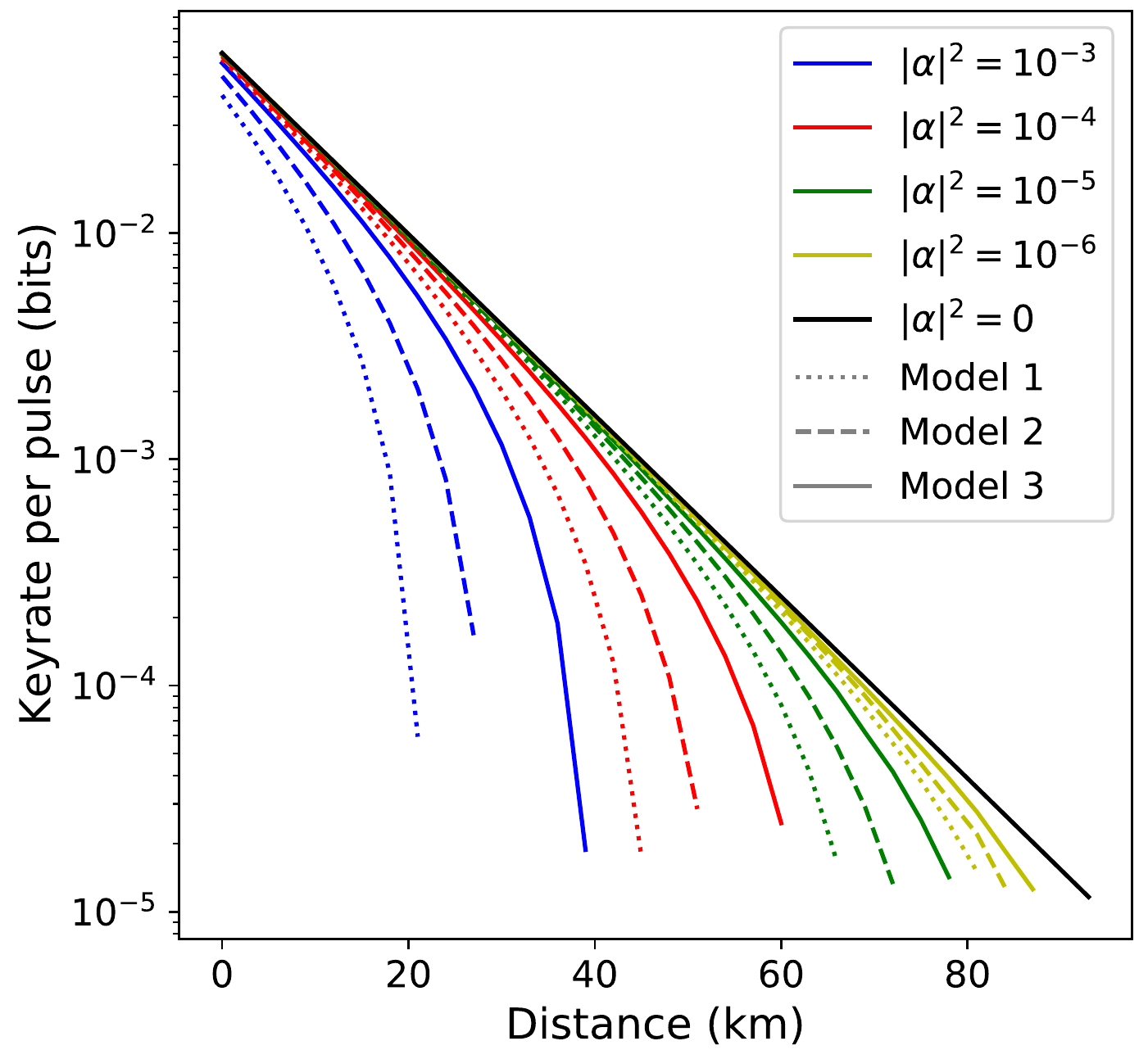}}
    \caption{Secret key rate as a function of Alice-Charlie distance for three different models of the leakage light. Model 1 corresponds to treating the leakage light as a superposition of vacuum (with amplitude $e^{-|\alpha|^2/2}$, same as a coherent state) and a state which leaks full encoding information. Model 2 corresponds to treating the leakage light as a coherent state $\ket{\alpha_0\cos\theta_x^i}_H\otimes\ket{\alpha_0\sin\theta_x^i e^{i\phi_x^i}}_V$ with the same polarization encoding parametrized by $\theta$ and $\phi$ as the signal state. Model 3 corresponds to treating the leakage light as a coherent state with total intensity $|\alpha|^2$, but with a time-dependent polarization, as given in Eq. \eqref{eq:model3}. Across all models, $|\alpha|^2$ can be interpreted as the intensity of the leakage signal. For each model, we plot the key rate for various values of $|\alpha|^2$ which we indicate with different colours.  In (a) we assume a decoy state protocol is used to characterize the single photon detection events, while in (b) we assume that the encoded modes of the signal state are perfect single photons.}
    \label{fig:leakage_model}
\end{figure}

Our main interest is determining how the key rate is affected by the presence of the side-channel. Here we investigate how the key rate changes depending on the model for the side-channel light, the intensity of the light, and on whether the encoded modes are sent as perfect single-photons source or as phase-randomized WCPs. In these simulations, we assume that Alice and Bob prepare BB84 states $\frac{\ket{H}\pm\ket{V}}{\sqrt{2}}$ and $\frac{\ket{H}\pm i\ket{V}}{\sqrt{2}}$, i.e. there are no encoding flaws. These simulations serve as a test to see how robust the key rate calculation is to the model of the leakage light; unsurprisingly, the key rate is highly dependent on the state of the leakage light.

In Fig. \ref{fig:leakage_model} (a), we plot the key rate as a function of Alice-Charlie distance, assuming a decoy state protocol, for the three models of leakage light. Additionally, we vary the intensity of the leakage light across several orders of magnitude; using the lowest intensity signals from \cite{Zhong2019} as an order-of-magnitude reference for highly attenuated light, this places realistic leakage light intensity somewhere on the order of $10^{-6}$ to $10^{-4}$. From this plot, we see that the most significant boosts in key rate come from a hardware solution of minimizing the intensity of the light in the side-channel; an order of magnitude reduction in intensity provides a greater improvement than refining the model of the leakage light state. However, there will likely always be some level of leakage light present between pulses. To mitigate this effect, it can be beneficial to carefully characterize the state of the side-channel. We see an improvement in the key rate when moving from Models 1 through 3, in that order. This reflects the intuition that the non-vacuum components of the states in the these models carry diminishing levels of information about the basis and bit choices. In MDI QKD, we require that Alice and Bob have complete characterization of their sources but no characterization of the detectors; thus, if they know the state of the side-channel (or at the very least the pairwise inner products of all the initial states), it is straightforward to incorporate more information about the state by modifying the inner product constraints (a simple software solution), rather than making pessimistic assumptions about the leakage light, as in Model 1, resulting in lower key rates.

Since the model for and intensity of the side-channel light has no bearing on the observed detection statistics, the detection constraints used to produce all the key rate curves in the figure are the same. As extra confirmation that the observed improvements in the key rate due to changing the model of the side-channel and the intensity of the side-channel light are independent of the observed detection statistics, in Fig. \ref{fig:leakage_model} (b), we provide the same key rate calculations, but assume a single-photon source for the encoded mode. We notice qualitatively the exact same trends as when using the decoy method, as expected.

\subsection{Sending Seemingly Redundant States Helps}\label{subsec:3vs4}
\begin{figure*}
    \centering
    \includegraphics[width=\textwidth]{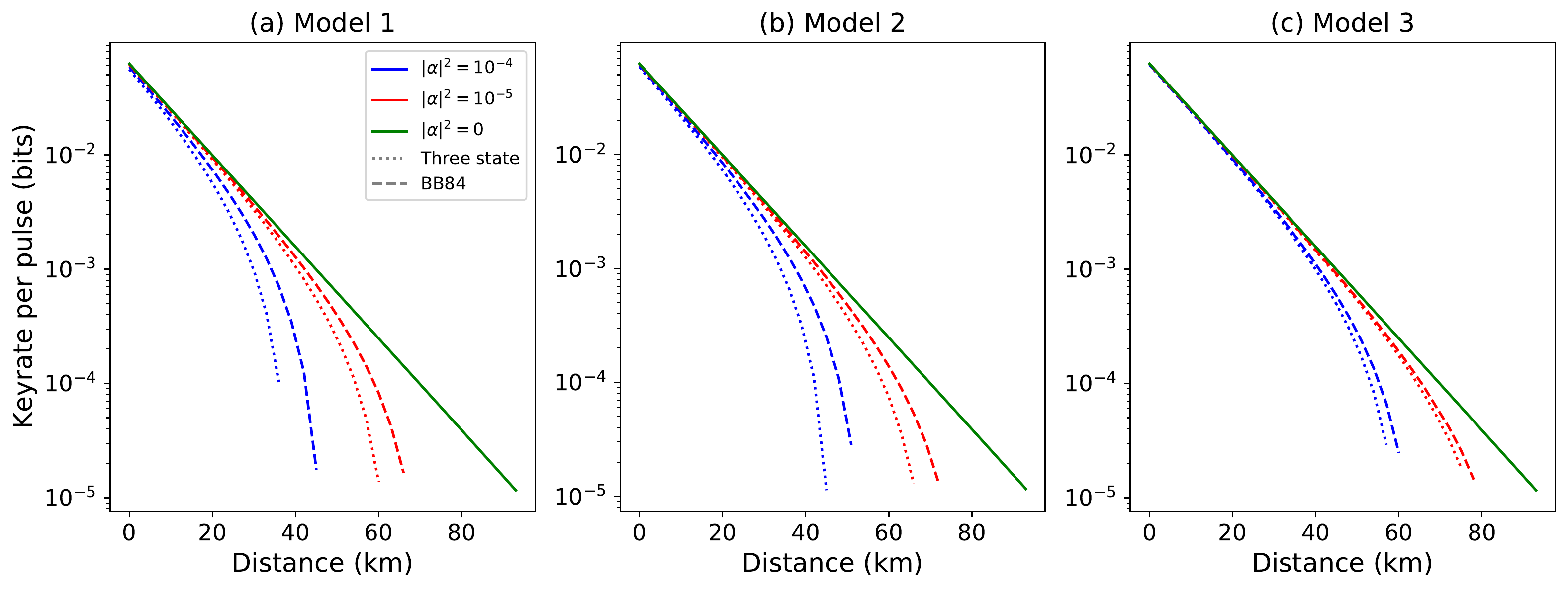}
    \caption{Secret key rate as a function of Alice-Charlie distance assuming a single photon source for the encoded mode states. Here we see the advantage of sending the four BB84 states instead of using the three-state protocol. This trend is true for different orders of magnitude of $|\alpha|^2$ and across all leakage state models, depicted in (a)-(c). This in contrast to the ideal case of $|\alpha|^2=0$, where BB84 and the three-state protocol yield the same key rates.}
    \label{fig:3vs4_state_singles}
\end{figure*}

\begin{figure*}
    \centering
    \includegraphics[width=\textwidth]{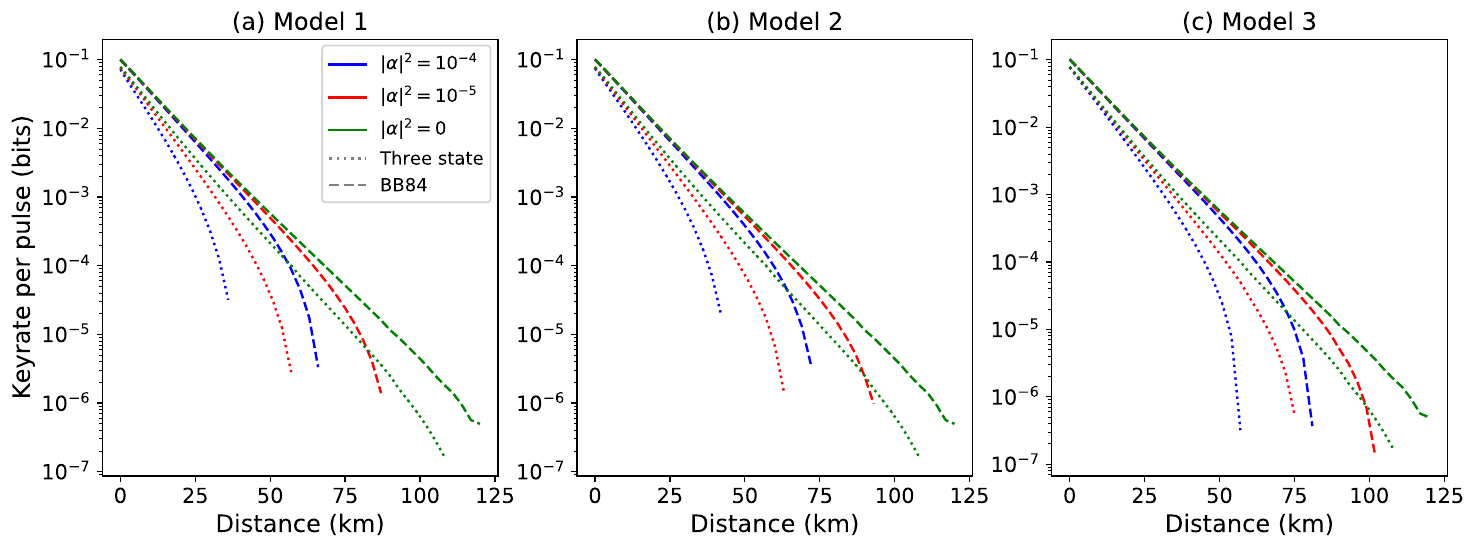}
    \caption{Secret key rate as a function of Alice-Charlie distance assuming a phase-randomized WCP and decoy state method. Like in Fig. \ref{fig:3vs4_state_singles}, we see the advantage of sending the four BB84 states instead of using the three-state protocol. Even in the ideal case of $|\alpha|^2=0$, BB84 outperforms the three-state protocol, since the detection probabilities only offer inequality constraints, meaning the extra fourth state does offer extra constraint to increase the key rate. When the side-channel is present, we also see that adding an extra state in BB84 can go so far as to achieve a higher key rate than using the three-state protocol with an order of magnitude lower leakage light intensity.}
    \label{fig:3vs4_state_decoys}
\end{figure*}

It is known that the with only three out of the four BB84 states and using all the detection statistics, that one can produce the same key rate as using all the BB84 states \cite{loss_tol}. Here, we are interested in whether the same conclusion extends to the case of when the source has a side-channel. While the fourth BB84 state is redundant in the case that there is no leakage light, when a side-channel is present, the extra state can help characterize Eve's attack on the leakage light. We certainly would not expect the key rate to decrease by sending an extra state, as the extra state will only provide additional inner product and detection constraints to those already provided by the other three states. 

In Fig. \ref{fig:3vs4_state_singles}, we plot the key rate for a single photon source, examining all three models of leakage light, and a couple different intensities. Across all models and intensities (except for $|\alpha|^2=0$) there is an increase in the key rate when all four BB84 states are used as opposed to only three. In Fig. \ref{fig:3vs4_state_decoys}, we plot the key rates again, this time assuming a decoy state protocol. In this case, the divergence between using three or four states is even more pronounced. Even the $|\alpha|^2=0$ case observes a boost in the key rate, since the detection statistic constraints in Eq. \eqref{eq:ineq_det_cons} are inequalities when using the decoy state method, so the extra detection statistics from the fourth state are useful in this case. The key rate boost achieved from switching from three to four states is so pronounced that it can even do better than decreasing the intensity of the leakage light: using four states with a leakage light intensity of $10^{-4}$ provides a higher key rate than using three states with a leakage intensity of $10^{-5}$.

The takeaway message from this analysis is that the three-state protocol is not as practical as BB84 in the presence of source side-channels. The extra resource savings of only having to use three states is undone by the loss of useful constraints that increase the key rate. We also simulated sending five and six states in the same plane of the Bloch sphere as the BB84 states to see if this provided even better key rates, but the key rate appeared to saturate with sending four states. Certainly sending additional states outside of this plane would increase the key rate, as expected from the six-state \cite{lo_sixstate} or tilted four state protocols \cite{loss_tol}, but this would require additional polarization modulation in the source, whereas it is easier to only vary the angle along one great circle of the Bloch sphere.

\section{Conclusion}

In this work, we have examined the problem of source side-channels in MDI QKD. We reviewed the decoy state method and a recent, versatile proof technique based on semidefinite programming which allows for information about the state of the side-channel to be incorporated into the key rate calculation. With this in hand, we examined a case example of a common MDI QKD source that employs a Faraday mirror for polarization stabilization. For this source, we identified a non-trivial, time-dependent side-channel due to leakage light between encoded optical pulses, provided a quantum optical model of the output, and linked the components of the source model to the security proof techniques. We then examined multiple protocol scenarios to understand strategies for improving the secret key rate under practical circumstances. 

We identified how the key rate calculation is affected by the information provided as constraints to the security proof. Most importantly, we saw how the model for the state of the leakage light can significantly impact the key rate, reaffirming that in MDI QKD security is derived in part from knowledge of the initial states sent by Alice and Bob, including any side-channel states. It is clear from our results that in practice one must carefully characterize side-channels, the reward of this work being higher key rates that come from not needing to take overly pessimistic assumptions of how much information is being leaked to the eavesdropper. On top of the importance of using the best available model for the side-channel, we found that in the presence of state-preparation flaws, Alice and Bob benefit from using all information at their disposal for the key rate calculation, i.e. all detection statistics, and all initial state information, rather than discarding cases when their basis choices do not match.

Having models for the state of the leakage light allowed us to develop concrete hardware strategies for mitigating the presence of the side-channel. Besides the obvious hardware improvement of simply suppressing the leakage light, two other physically implementable strategies emerged for when leakage light is present: first, although the three-state protocol promises the same key rates as BB84, when leakage light is present, Alice and Bob can get better key rates by sending all four BB84 states, as the statistics from the normally redundant fourth state actually help to better constrain Eve's attack on the side-channel. Second, while the choice of which test state to send from the Bloch sphere typically does not matter, here we find that in the presence of leakage light, some test states provide better key rates than others, indicating the advantage of optimizing which states to send as a function of distance. Even though we considered a representative case example, we expect that the strategies we developed to mitigate the side channel to be widely applicable to other leaky sources.

While this work examined strategies for treating source side-channels in MDI QKD, the source we considered had the advantage of not leaking information about the intensity setting choice and random phase value of the decoy state protocol, meaning we were able to use the decoy state method with only modifications to how we interpret the output of the linear programs in the security proof. An open problem is how to simultaneously mitigate more general source side-channels that leak information about both the encoding information, as we investigated, and the decoy state method intensity and phase parameters. It would be worthwhile to investigate merging the analysis presented here with the proof technique from \cite{wang2021measurement} for treating intensity and phase information leakage.

\begin{acknowledgments}
We are grateful for discussions with Thomas Van Himbeeck, Ignatius William Primaatmaja and Emilien Lavie, and especially thankful to Wenyuan Wang for extensive discussions about decoy states. J.E.B is supported by an Ontario Graduate Scholarship. We thank financial support from NSERC, MITACS, CFI, ORF, Royal Bank of Canada, Huawei Technology Canada, CRCEF and the University of Hong Kong start-up funding.
\end{acknowledgments}

\appendix
\section{Review of the Decoy State Method for MDI QKD}\label{app:decoy_lin_prog}

When Alice and Bob encode their secret key in a single photon degree of freedom, photon number splitting attacks are a method for an eavesdropper to exploit multiphoton output of the optical source to learn the secret key \cite{GLLP2003}. As a consequence, only single photon detection events are usable to characterize the amount of information the eavesdropper has about the key; however, multiphoton events can still be used to characterize the correctness of the key. The decoy state method allows Alice and Bob to characterize the photon number statistics of the eavesdropper-controlled channel, and in turn bound security based on the detection events that arose from the single photon components of the source's optical output \cite{Lo2005}.

Practically, the decoy state method for MDI QKD with polarization encoding consists of Alice and Bob each preparing phase-randomized weak coherent pulses (WCPs) with varying intensities \cite{Ma2012_mdi_decoy}. Each pulse is polarized according to the protocol, e.g. BB84 \cite{Lo2012} or three-state \cite{loss_tol}. In this case, we can write the photon number distribution of Alice and Bob's states as:
\begin{equation}\label{eq:decoy_photon_dist}
    p(m,n|k,l) = \frac{e^{-(\mu_k+\nu_l)}\mu_k^m\nu_l^n}{m!n!},
\end{equation}
where $k$ ($l$) refers to Alice's (Bob's) optical intensity setting $\mu_k$ ($\nu_l$). This assumes that the intensity settings are completely independent of the polarization basis and bit setting choices. If they each use $N$ intensity settings (typically three is sufficient), then for given basis $(i,j)$ and bit $(x,y)$ choices, they have $N^2$ linear equations for the detection probabilities as a function of photon number:
\begin{equation}\label{eq:tot_det_prob}
    Q^{i,j,x,y}_{k,l}=\sum_{m,n} \frac{e^{-(\mu_k+\nu_l)}\mu_k^m\nu_l^n}{m!n!} p_{\texttt{pass},m,n}^{i,j,x,y}.
\end{equation}
Here, $Q^{i,j,x,y}_{k,l}$ is the observed probability of Eve announcing that a round passed given that Alice and Bob chose intensity settings $(k,l)$ along with basis and bit choices $(i,j,x,y)$. $p_{\texttt{pass},m,n}^{i,j,x,y}$ is the probability that the round passes due Alice (Bob) sending $m$ ($n$) photons, and given basis and bit choices $(i,j,x,y)$. Since we assume that the intensity setting choices are independent of the basis and bit choices, and that phase randomization of the WCP is perfect, $p_{\texttt{pass},m,n}^{i,j,x,y}$ is independent of $(k,l)$. These $N^2$ linear equations can be used in a linear program to determine upper and lower bounds on all the detection probabilities due to single photon components of the optical output $p_{\texttt{pass},1,1}^{i,j,x,y}$. This allows us to bound the relevant detection statistics for computing security.

To construct a linear program to calculate upper and lower bounds on the single photon detection probabilities of a decoy state MDI QKD protocol \cite{Ma2012_mdi_decoy}, we first identify the variables of the optimization as $p_{\texttt{pass},m,n}^{i,j,x,y}$. To establish a lower (upper) bound on $p_{\texttt{pass},1,1}^{i,j,x,y}$ given these constraints, we solve the linear program to find the minimum (maximum) possible value of $p_{\texttt{pass},1,1}^{i,j,x,y}$ consistent with the constraints. If Alice and Bob each have $n_A$ and $n_B$ basis choice settings, each basis choice associated with two bit choices, we repeat the process of finding lower and upper bounds for all $n_A \times n_B\times2\times 2$ combinations of $(i,j,x,y)$.

Since there are in principle infinitely many $p_{\texttt{pass},m,n}^{i,j,x,y}$, for a practical linear program, we impose a cutoff photon number $N_{\max}$. In that case, the $N^2$ linear equality constraints become $2N^2$ linear inequality constraints. The first $N^2$ constraints are:
\begin{equation}
    Q^{i,j,x,y}_{k,l} \geq \sum_{m,n=0}^{N_{\max}} \frac{e^{-(\mu_k+\nu_l)}\mu_k^m\nu_l^n}{m!n!} p_{\texttt{pass},m,n}^{i,j,x,y},
\end{equation}
stemming from the fact that summing up to the cutoff will yield a value less than the total detection probability. For the next $N^2$ constraints, we find:
\begin{equation}
\begin{split}
       &1 - \sum_{m,n=0}^{N_{\max}} \frac{e^{-(\mu_k+\nu_l)}\mu_k^m\nu_l^n}{m!n!}\\
     &\geq\sum_{m,n=N_{\max}+1}^{\infty} \frac{e^{-(\mu_k+\nu_l)}\mu_k^m\nu_l^n}{m!n!} p_{\texttt{pass},m,n}^{i,j,x,y}.  
\end{split}
\end{equation}
which means we can provide the constraints:
\begin{equation}
\begin{split}
    &\sum_{m,n=0}^{N_{\max}} \frac{e^{-(\mu_k+\nu_l)}\mu_k^m\nu_l^n}{m!n!} p_{\texttt{pass},m,n}^{i,j,x,y}\\
    &\geq Q^{i,j,x,y}_{k,l}+\sum_{m,n=0}^{N_{\max}} \frac{e^{-(\mu_k+\nu_l)}\mu_k^m\nu_l^n}{m!n!} -1.    
\end{split}
\end{equation}
In practice for our calculations in the main text, we found an $N_{max}$ of 10 photons was sufficient to provide good upper and lower bounds on $p_{\texttt{pass},1,1}^{i,j,x,y}$ while not taking too long to compute.

Using the decoy method, a lower bound on the secret key rate in an MDI QKD protocol is provided by \cite{Ma2012_mdi_decoy}:
\begin{equation}\label{eq:KR_formula}
    R \geq  p_{\texttt{pass},1,1}^{0,0}[1-h_2(e_{ph,1,1})] - Q^{0,0}_{N,N} h_2(E_{bit}),
\end{equation}
where $h_2(\cdot)$ is the binary entropy function. $Q^{0,0}_{N,N}$ is the detection probability of outcomes that generate raw key, and is given by:
\begin{equation}
    Q^{0,0}_{N,N} = \sum_{x,y} Q^{0,0,x,y}_{N,N}
\end{equation}
where we choose, without loss of generality, $(i,j)=(0,0)$ to be the key generation basis and $(k,l)=(N,N)$ to be the key generation intensities. $E_{bit}$ is the bit error rate of the raw key, given by:
\begin{equation}
    E_{bit} = \frac{\sum_{x\neq y} Q^{0,0,x,y}_{N,N}}{Q^{0,0}_{N,N}}
\end{equation}
$p_{\texttt{pass},1,1}^{0,0}$ is the detection probability due to the single photon components of Alice and Bob's optical output:
\begin{equation}
    p_{\texttt{pass},1,1}^{0,0} = \sum_{x,y} p_{\texttt{pass},1,1}^{0,0,x,y}.
\end{equation}
Finally, $e_{ph,1,1}$ is the phase error rate of the protocol which we will more precisely define in the next section. Briefly, were we to consider a virtual picture of the protocol in which the single photon components of the optical output are entangled with qubits kept in Alice and Bob's labs, the phase error is the probability those qubits end up in a target Bell state up to a phase error. Like the single photon detection probabilities, it is not a directly observable quantity of the protocol and must be bounded. Were Alice and Bob to be able to perfectly prepare eigenstates of the conjugate basis to the key generation basis, then $e_{ph,1,1}$ can also be bounded with a simple linear program \cite{Ma2012_mdi_decoy}. When the sources have preparation flaws amounting to constant polarization offsets, a series of linear programs are required (see Appendix A of \cite{LT_exp}). In the case that the sources have preparation flaws and have a side-channel, we can employ the more recent technique \cite{Primaatmaja2019} for bounding the phase error rate that employs semidefinite programming; that technique is reviewed in Section \ref{subsec:proof}.

\section{Comparison to the proof technique from \texorpdfstring{\cite{Pereira2019}}{TEXT}}\label{app:pereira_comp}

Here we will compare the proof technique we are using to the technique from \cite{Pereira2019}. We will show that \cite{Pereira2019} relaxes the SDP inherent to optimizing the phase error rate to a linear program, which we would expect to give equal or lower bounds on the key rate than computing the full SDP. For simplicity, we will consider a protocol where when Alice and Bob choose the Z basis, they perfectly prepare qubit states $\ket{0},\ket{1}$, but we allow for their test states to have leakage components outside of the qubit subspace spanned by $\{\ket{0},\ket{1}\}$. The following comparison can be generalized in a straightforward manner to arbitrary initial states.

Let $U$ be the unitary that takes $\ket{\psi^{i}_{x}\phi^j_y}_{A,B}\rightarrow \sum_{z} \ket{e_{x,y,z}^{i,j}}$. Thus, for this case of initial states, the phase error rate would be given by:
\begin{equation}
    e_{ph} = \frac{Tr\left[\ket{P}\bra{P}_Z U \left(\frac{\id+\sigma_X\otimes\sigma_X}{2}\right)_{A,B} U^\dagger\right]}{\sum_{x,y}p_{\texttt{pass}}^{0,0,x,y}},
\end{equation}
where $\sigma_m$, $m=I,X,Y,Z$ refer to Pauli operators in the qubit space spanned by $\{\ket{0},\ket{1}\}$. Following \cite{loss_tol,Pereira2019}, $e_{ph}$ can be decomposed in terms of the transmission rates of the Pauli operators $q_{\texttt{pass}|i,j}=Tr\left(\ket{P}\bra{P}_Z U \sigma_i\otimes\sigma_j U^\dagger\right)$, since the Pauli operators form a basis for any operator. Were $\epsilon=1$, then we could use the states Alice and Bob send to exactly solve for $q_{\texttt{pass}|i,j}$ (assuming their test states are some superposition of $\{\ket{0},\ket{1}\}$). However, when their signal states have a leakage component, we cannot exactly constrain these quantities; in \cite{Pereira2019} $q_{\texttt{pass}|i,j}$ form the variables of a linear program that are optimized to determine a lower bound on $e_{ph}$.

First, we write Alice and Bob's signal states as linear combinations of states in a two-qubit space, and a space orthogonal to it (the leakage space), just as in Eq. (1) of \cite{Pereira2019}:
\begin{equation}\label{eq:qubit_space}
    \ket{\psi^{i}_{x}\phi^j_y}_{A,B} = a^{i,j}_{x,y}\ket{\tilde{\psi}^{i}_{x}\tilde{\phi}^{j}_{y}}_{A,B} + {b^{i,j}_{x,y}}\ket{{\tilde{\psi}^{i}_{x}\tilde{\phi}^{j}_{y}}^{\perp}}_{A,B}
\end{equation}
where the orthogonality of the two-qubit and leakage space means $\braket{\cdot|\cdot^\perp}_{A,B}=0$.

Next, the detection probabilities provide the constraint:
\begin{equation}
    p_{\texttt{pass}}^{i,j,x,y} 
    = Tr\left(\ket{P}\bra{P}_Z U \ket{\psi^{i}_{x}\phi^j_y}\bra{\psi^{i}_{x}\phi^j_y}_{A,B} U^\dagger\right).
\end{equation}
Using the decomposition from Eq. \eqref{eq:qubit_space}, we find this is also equal to:
\begin{widetext}
\begin{equation}\label{eq:pdet_pereira}
\begin{split}
    p_{\texttt{pass}}^{i,j,x,y} 
    =&|a^{i,j}_{x,y}|^2 Tr\left[\ket{P}\bra{P}_Z U 
     \ket{\tilde{\psi}^{i}_{x}\tilde{\phi}^{j}_{y}}\bra{\tilde{\psi}^{i}_{x}\tilde{\phi}^{j}_{y}}_{A,B} U^\dagger\right]
    + Tr\left[\ket{P}\bra{P}_Z U \left(a^{i,j}_{x,y}{b^{i,j}_{x,y}}^*\ket{\tilde{\psi}^{i}_{x}\tilde{\phi}^{j}_{y}}\bra{{\tilde{\psi}^{i}_{x}\tilde{\phi}^{j}_{y}}^{\perp}}\right.\right.\\
    &\hspace{2.5cm}\left.\left.+{a^{i,j}_{x,y}}^*{b^{i,j}_{x,y}}\ket{{\tilde{\psi}^{i}_{x}\tilde{\phi}^{j}_{y}}^{\perp}}\bra{\tilde{\psi}^{i}_{x}\tilde{\phi}^{j}_{y}}
    +|b^{i,j}_{x,y}|^2\ket{{\tilde{\psi}^{i}_{x}\tilde{\phi}^{j}_{y}}^{\perp}}\bra{{\tilde{\psi}^{i}_{x}\tilde{\phi}^{j}_{y}}^{\perp}}
    \right)_{A,B} U^\dagger\right]
\end{split}
\end{equation}
\end{widetext}
which coincides with the MDI QKD version of Eq. (19) from \cite{Pereira2019}. Since the operator $\ket{\tilde{\psi}^{i}_{x}\tilde{\phi}^{j}_{y}}\bra{\tilde{\psi}^{i}_{x}\tilde{\phi}^{j}_{y}}$ lives in the two-qubit subspace, it too can be written in terms of the Pauli operators $\sigma_{m}\otimes\sigma_{n}$, meaning the first trace term in Eq. \eqref{eq:pdet_pereira} can be written in terms of $q_{\texttt{pass}|i,j}$. Because of the second trace term, we cannot solve for them exactly as is done in the loss-tolerant proof technique \cite{loss_tol}.

Rather than solving the semidefinite program for $e_{ph}$ using the linear equality constraints provided by the detection probabilities, \cite{Pereira2019} considers a relaxation to a linear program. Specifically, the second trace term in Eq. \eqref{eq:pdet_pereira} can be bounded above and below by the maximum and minimum eigenvalues of the matrix:
\begin{equation}
    M^{i,j}_{x,y} = \begin{pmatrix}
        0 & a^{i,j}_{x,y}{b^{i,j}_{x,y}}^*\\
        {a^{i,j}_{x,y}}^*{b^{i,j}_{x,y}} & |b^{i,j}_{x,y}|^2
        \end{pmatrix}.
\end{equation}
Thus, \ref{eq:pdet_pereira} leads to inequalities linear in $q_{\texttt{pass}|i,j}$, which can be used as constraints in a linear program to find an upper bound for $e_{ph}$. However, because the exact equality constraints coming from the detection probabilities have been relaxed using the maximum and minimum eigenvalues of $M^{i,j}_{x,y}$, we would expect that this would lead to a greater upper bound on $e_{ph}$ (and hence a weaker lower bound on the key rate) than if the exact constraints were kept, as they would be in the numerical approach from \cite{Primaatmaja2019} that we reviewed in Section \ref{subsec:proof}. 

Note that the proof approach we have used in this paper does away with needing to frame the calculation of $e_{ph}$ in terms of $q_{\texttt{pass}|i,j}$ (even though we could, in principle, do so since they are linear functions of the elements of Eve's Gram matrix); after all, since the signal states are no longer qubits, we need not make the distinction between a qubit subspace and the leakage space, since Eve's operation can blend these two spaces. Instead, given that the phase error can be expressed in terms of the elements of a positive semidefinite matrix associated with Eve's information, and given that we have linear equality constraints on this matrix, the phase error can be maximized directly with a simple SDP, rather than relaxing to a linear program.

\begin{figure}
    \includegraphics[width=\columnwidth]{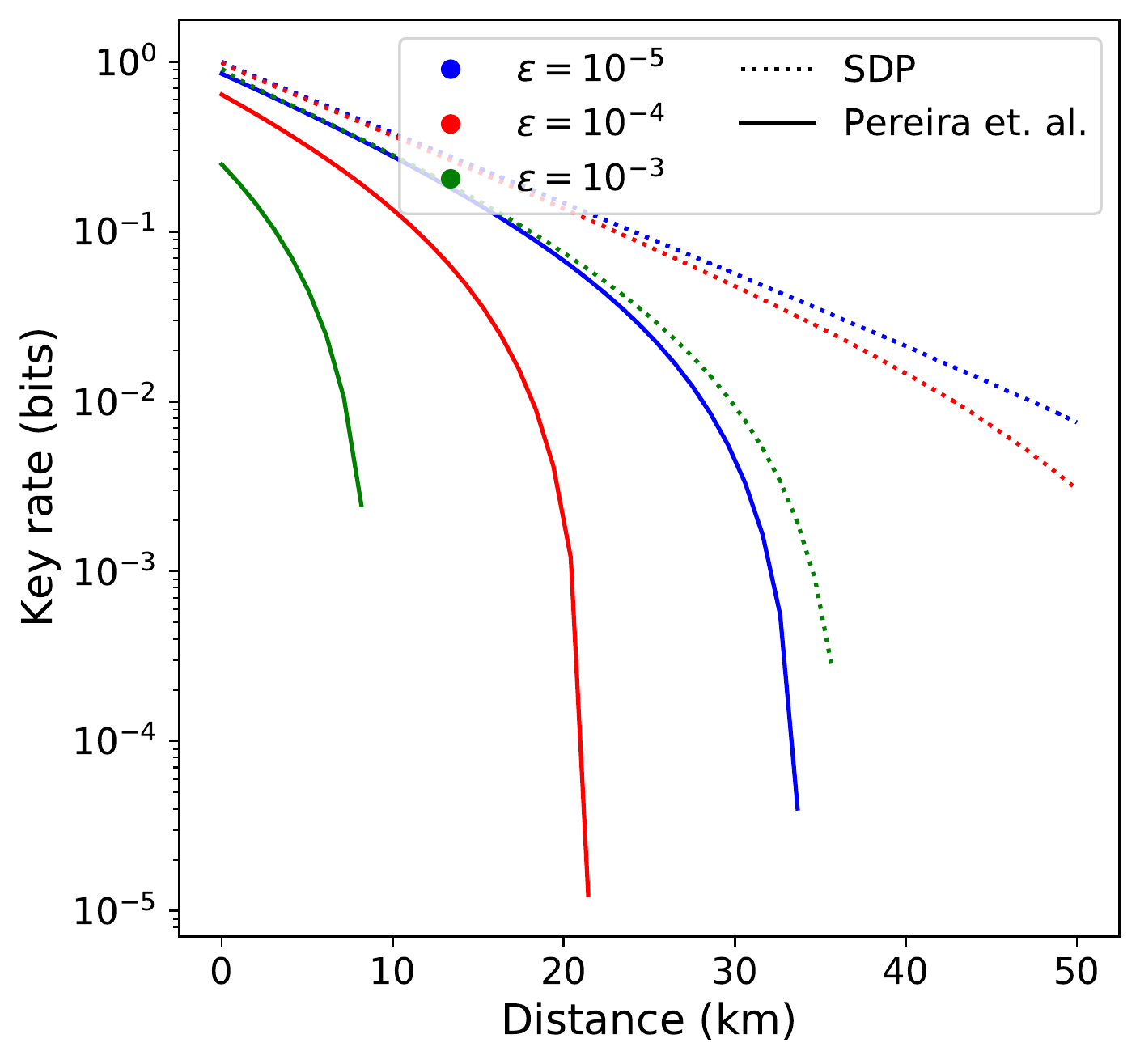}
    \caption{Key rate vs. distance for various values of $\epsilon$, calculated using the SDP method reviewed in \ref{subsec:proof} and the method from \cite{Pereira2019}. We see the advantage of the SDP approach over the relaxation to a linear program, as done in \cite{Pereira2019}.}
    \label{fig:sdp_vs_pereira}
\end{figure}

As an example to demonstrate the superiority of the SDP method over the method from \cite{Pereira2019}, we consider a toy example of the three state protocol with a single photon source, for which Alice and Bob prepare a leaky third state, 
\begin{equation}
    \ket{+}_{enc}(\sqrt{\epsilon}\ket{vac}_{leak}+\sqrt{1-\epsilon}_{enc}\ket{1}_{leak}),
\end{equation}
as opposed to the ideal $\ket{+}_{enc}$. We assume a detection efficiency of 1 and a dark count rate of $10^{-6}$. In Fig. \ref{fig:sdp_vs_pereira}, we plot the key rates calculated using the SDP method we reviewed in Section \ref{subsec:proof} and using the method from \cite{Pereira2019}. We find that across values of $\epsilon$, the SDP method performs much better.\\

\section{Derivation of Figure \ref{fig:leakage_pol}}\label{app:fig_derivation}

Here, we derive the fractional phase change applied to the leakage light as a function of time, as shown in Figure \ref{fig:leakage_pol}.

\begin{figure*}
    \centering
    \includegraphics[width=0.7\textwidth]{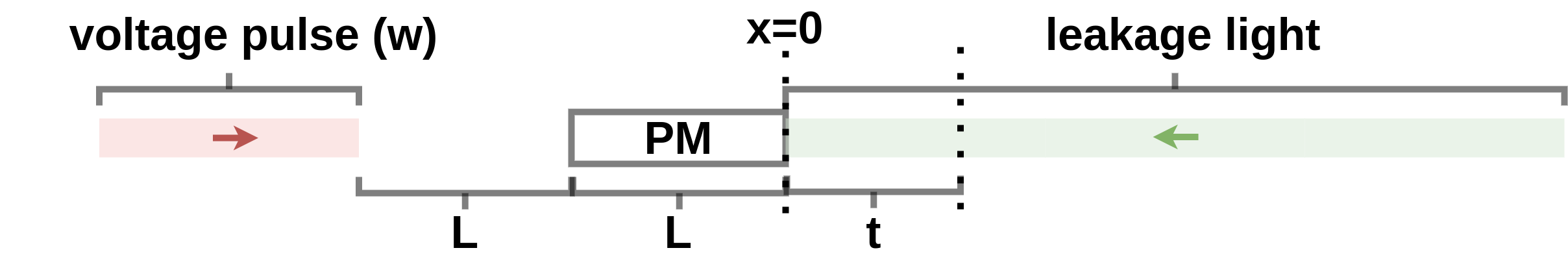}
    \caption{An illustration depicting a moment in time as a voltage pulse approaches the phase modulator (PM).}
    \label{fig:appB_2}
\end{figure*}

\begin{figure*}
    \centering
    \includegraphics[width=\textwidth]{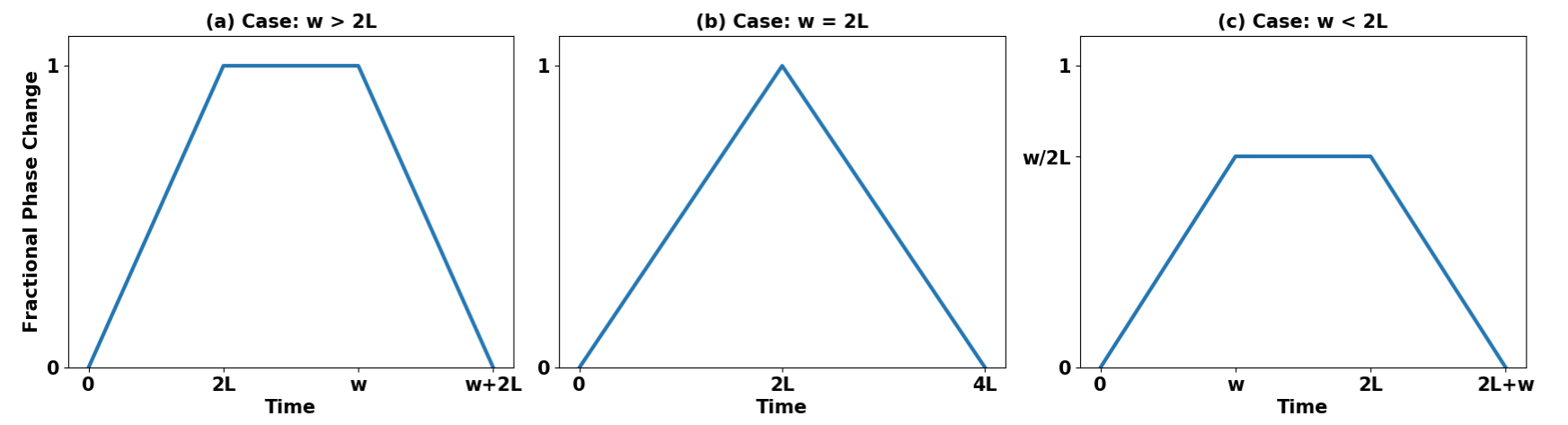}
    \caption{The various solutions for the integral in Eq.  \ref{eq:fractional_phase_change}.}
    \label{fig:appB_3}
\end{figure*}

Refer to the experimental setup shown in Figure \ref{fig:exp_setup}, specifically the polarization modulation unit. First, optical pulses travel forward through the PM for the purpose of polarization modulation. Simultaneously, voltage pulses overlapping in time with the optical pulses are sent into the PM, propagating in the same direction as the optical pulses. The PM is designed such that the optical and voltage pulses travel through the PM at the same speed. The voltage is what enables a phase change and therefore a polarization change. Since the leakage light is not meant to encode information, voltage is not sent through the PM as this light travels through the first time. 

However, when traveling back through the PM after reflection from the Faraday mirror, the leakage light will inevitably collide temporally with a voltage pulse that is travelling in the opposite direction along with an optical pulse it is intended to modulate. 

The overall phase modulation experienced by a temporal slice of light after traveling through the PM can be expressed as:
\begin{equation}\label{eq:B1}
    \phi = K\int_{0}^{L} V(z) dz.
\end{equation} 
Here, $L$ represents the length of the PM and $V(z)$ represents the applied voltage overlapping with the slice of light. $K$ is simply a proportionality constant. When light is travelling in the same direction as the voltage wave through the phase modulator, Eq. \eqref{eq:B1} reduces to $K \times V \times L$. This occurs due to the voltage, which is moving at the same speed as the light, being a constant along the length of the PM. 

In our case, we are also dealing with leakage light that is travelling in the opposite direction. We will use $L$ to refer to the PM length and $w$ to refer to the width of the square voltage pulses sent to the PM. Given these parameters, we can determine the phase change experienced by the leakage light as follows: 
\begin{enumerate}

    \item  Refer to Figure \ref{fig:appB_2}. We will use this moment in time as our starting point. First we will create a coordinate system by defining $x=0$ to be the right hand edge of the phase modulator. We can parametrize a slice of leakage light with $t$, the time it crosses the point $x=0$. 
    \item At the point in time shown in Figure \ref{fig:appB_2}, we can define the voltage pulse as $A(x)=H(x+2L+w)-H(x+2L)$ (a square pulse) and the leakage light as $B(x)=H(x)$ where $H$ refers to the Heaviside step function. 
    \item Now, notice that the movement in time of the voltage pulse and leakage light can also be incorporated into these functions. After $\tau$ ps, the function defining the voltage pulse will become $A(x-\tau)$ while the function defining the leakage light will become $B(x+\tau)$.
    \item Notice that $A(x-\tau)\times B(x+\tau)$ represents the overlap between the voltage and leakage light at position $x$ and time $\tau$. It has a value of 1 if there is an overlap and a value of 0 if there is no overlap. 
    \item Now, suppose we want to calculate the amount of time for which the slice $t$ experiences an overlap within the phase modulator. We need to integrate $A(x-\tau)\times B(x+\tau)$ from $\tau=t$ to $\tau=t+L$. In other words, we need to integrate the overlap function over the values of $\tau$ for which the slice at position $t$ is inside the phase modulator. 
    \item The slice at position $t$ has an $x$ position of $t-\tau$ at time $\tau$. Substitute this into the integral for $x$.
    \item The resulting integral is as follows:
    \begin{equation}
        \int_{t}^{t+L} A(t-2\tau)\times B(t) \,d\tau.
    \end{equation} 
    This integral represents the amount of time the slice $t$ is in contact with a voltage pulse within the phase modulator. Recall that the optical pulse which is travelling along with (in the same direction as) this voltage pulse would be in contact along the entire length of the phase modulator ($L$). Therefore, the phase change experienced by slice $t$ is
    \begin{equation}
        \label{eq:fractional_phase_change}
        \frac{1}{L}\int_{t}^{t+L} A(t-2\tau)\times B(t) \,d\tau
    \end{equation}
    when written as a fraction of the phase change experienced by the pulse. The solution to this integral is shown in Figure \ref{fig:appB_3}. In our particular experimental setup, $L$ = 150 ps and $w$ = 200 ps. The value of the integral for these parameter values is plotted in Figure \ref{fig:leakage_pol}. The maximal fractional phase change is $\frac{2}{3}$. 
    
\end{enumerate}

\section{Additional key rate results}\label{app:more_KR}

Here, we present some additional exploration of strategies that can be used to increase the key rate in the presence of a source side-channel.

\subsection{Basis Mismatch Constraints}\label{subsec:mismatch}

\begin{figure}
    \centering
    \subfloat[Single-photon source]{\includegraphics[width=\columnwidth]{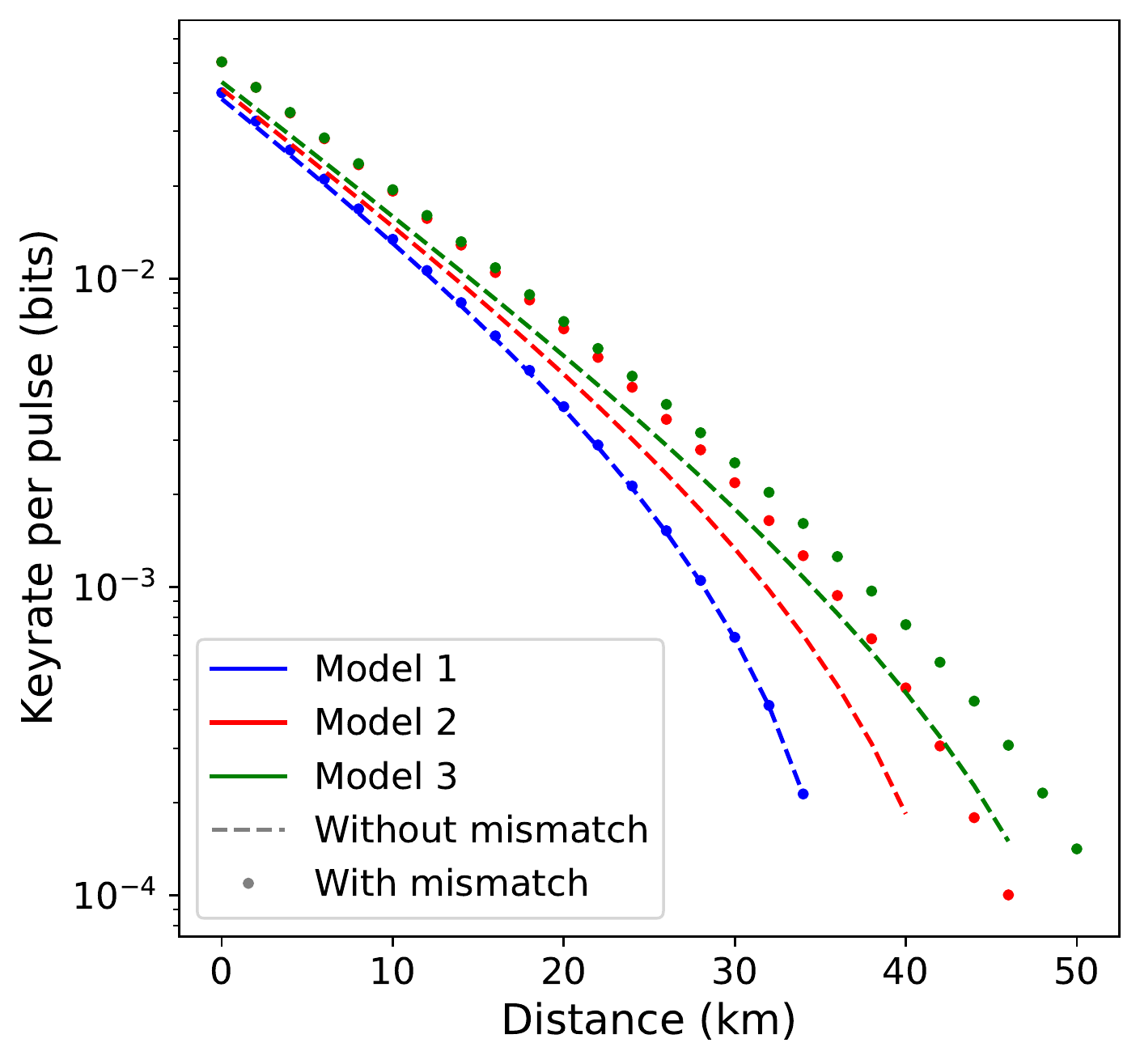}}
    \newline
    \subfloat[Decoy method]{\includegraphics[width=\columnwidth]{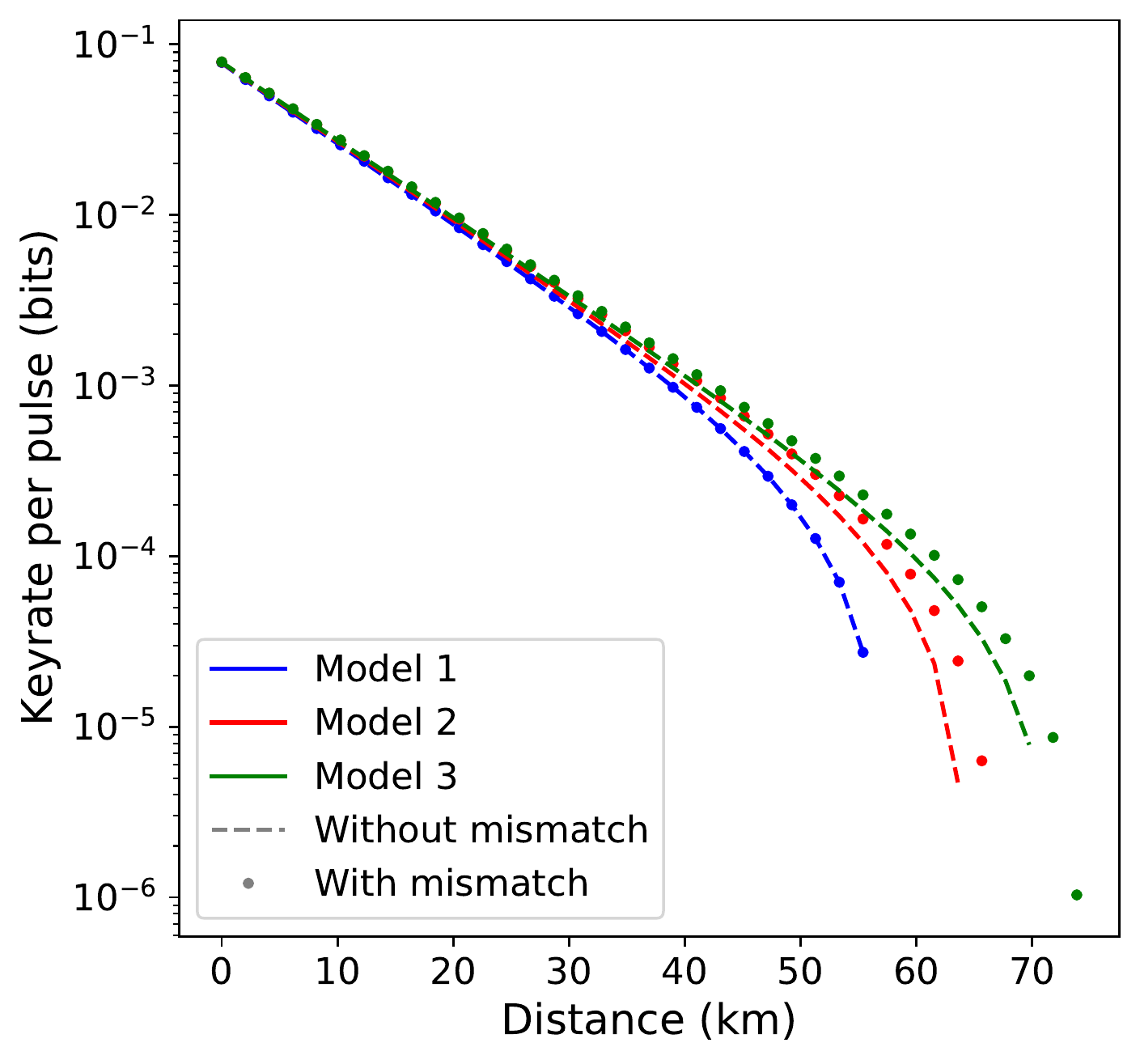}}
    \caption{Secret key rate as a function of Alice-Charlie distance for Models 1-3 of the leakage light. Here, we investigate whether using all initial state inner products and detection statistics, as opposed to just the cases when Alice and Bob choose the same basis, benefit the key rate. We consider the case of BB84 with a preparation flaw, and a side-channel with $|\alpha|^2=10^{-4}$, for both (a) a single-photon source, and (b) the decoy state method. For (b), Alice and Bob each use four decoy intensities. We observe that for Models 2 and 3, the key rate benefits from considering all inner products and detection statistics available.}
    \label{fig:mismatch_statistics}
\end{figure}

We are interested in knowing whether any advantage can be gained by using all the detection statistics and all the initial state inner products, including when Alice and Bob's bases do not match, as opposed to simply using the cases when the basis choices match $(i=j)$. We observed that when Alice and Bob prepare the BB84 states perfectly, using the basis mismatch statistics and inner products did not produce an increase in the key rate, even in the presence of a side-channel. With perfect state preparation, we know that the conjugate basis statistics alone are strict enough constraints to provide the phase error when there is no leakage light, and we confirm numerically that this extends to the case when leakage light is present.

However, we know that when Alice and Bob have a preparation flaw for their states, i.e. a constant offset angle on the Bloch sphere, the mismatch statistics can help better characterize the key rate \cite{loss_tol}. Since the Bloch sphere angle affects the associated side-channel state, the inner products, and the detection statistics, it is more difficult to predict how the key rate will respond to a preparation flaw, and whether using full or partial detection statistics in the SDP constraints benefits the key rate. For these simulations, we use the preparation flaw model from Appendix D of \cite{LT_exp}, with the Bloch sphere offset angle parameter $\delta=0.1$. In this case, we fix the leakage light intensity to $|\alpha|^2=10^{-4}$. 

In Fig. \ref{fig:mismatch_statistics} (a), we plot the key rate assuming a single-photon source for the encoded mode. For Model 1, we barely see any increase in the key rate when using full vs. partial detection statistics; this makes sense, since the non-vacuum component of the side-channel state leaks full encoding information, independent of Bloch sphere angle. For Models 2 and 3, we observe a boost in the key rate when using full detection statistics and inner products as constraints. This indicates that when one has a preparation flaw, and the side-channel state depends on the preparation flaw, it is best to use all information available from the detection statistics and initial state inner products.

In Fig. \ref{fig:mismatch_statistics} (b), we consider the same situation but with a decoy state protocol. For this scenario, we add a fourth decoy with vacuum intensity, and observe an increase in the key rate when using full detection statistics and initial inner products as constraints in Models 2 and 3. Like before, we do not observe an increase in the key rate for Model 1. When we only considered three decoy intensities, we did not observe a meaningful increase in the key rate, likely because the three decoy intensities did not allow for tight enough constraints on the single photon detection statistics, so adding more detection statistics as constraints did not help since the constraints were too loose.

\subsection{Choice of Test States Matters}\label{subsec:angles}
Another example of divergence between ideal sources and sources with side-channels occurs in the choice of which test states to send. In the ideal case, if Alice and Bob prepare two orthogonal polarization states, they need only send one other state to achieve the same key rate as BB84 \cite{loss_tol}; the location of that state on the Bloch sphere does not matter (as long as it is not the same state as the first two). Here we are interested to see whether this changes in the presence of a source side-channel.

To study this problem, we fix the channel distance, a leakage light intensity of $|\alpha|^2=10^{-4}$, fix Alice and Bob to send encoded single photon components $\frac{\ket{H}\pm\ket{V}}{\sqrt{2}}$ as two of their states, then vary the azimuthal angle of the other two states sent $\frac{\ket{H}\pm e^{i\phi}\ket{V}}{\sqrt{2}}$, and observe how the key rate changes. By symmetry, we need only vary $\phi\in[0,\pi]$. 

\begin{figure}
    \includegraphics[width=\columnwidth]{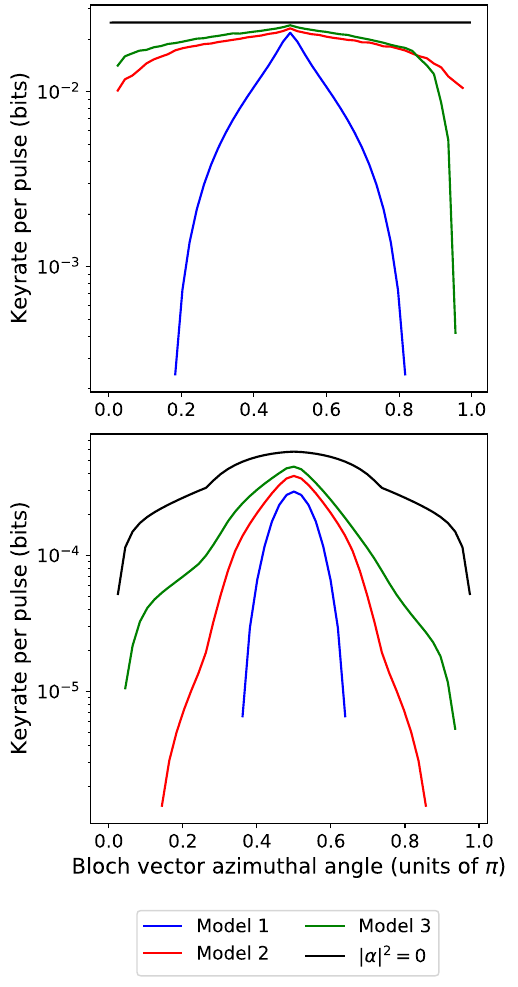}
    \caption{Key rate vs. azimuthal angle of the test states, for the case of no leakage light, and for leakage light with intensity $|\alpha|^2=10^{-4}$ treated with Models 1-3. The top (bottom) figure provides results for a single-photon source (a decoy state method) at a distance of 10 km (50 km). While the choice of test state is less relevant for the case of no leakage light, it can significantly decrease the key rate in the presence of leakage light, prompting the need to optimize which test states are used at a given distance.}
    \label{fig:test_state_angle}
\end{figure}

In top of Fig. \ref{fig:test_state_angle}, we plot the results assuming a single photon source and a distance of 10 km. As expected, the key rate is independent of $\phi$ when there is no leakage light. In the presence of leakage light, $\phi=\pi/2$ still remains as the optimum test state to send, but the key rate drops off away from that point, most dramatically for Model 1. Interestingly, there is even a region for which Model 2 outperforms Model 3. To explain this, we can go to the Gram matrix formed by the initial states which form the constraints on the RHS of Eq. \ref{eq:inner_prod}. If we calculate the trace distance between the Gram matrix of Model 2 and the Gram matrix created by the ideal qubit states $\{\frac{\ket{H}\pm\ket{V}}{\sqrt{2}},\frac{\ket{H}\pm e^{i\phi}\ket{V}}{\sqrt{2}}\}$ as a function of $\phi$, we find that it is symmetric about $\phi=\pi/2$; however, doing the same for the Gram matrix of Model 3, we find that the trace distance is not symmetric about that point due to the time-dependent nature of the underlying states and the way the inner product is calculated in Eq. \ref{eq:td_inner}. As $\phi$ increases, the Gram matrix of Model 3 eventually becomes a further distance from ideal than the Gram matrix of Model 2 for $\phi\gtrsim0.8\pi$, so it is conceivable the key rate for Model 3 can perform worse in that region. Of course, the key rate depends on much more than just this trace distance, since the angle also changes the constraints provided by the detection statistics, but this provides some intuition as to why Model 2 can outperform Model 3 in certain regimes.

In bottom of Fig. \ref{fig:test_state_angle}, we plot the key rates assuming a decoy state method and a distance of 50 km. Here, even the case of zero leakage light has some sensitivity to the angle of the test state. $\phi=\pi/2$ is still the optimal test state across all models. Like before, there is a limited range of $\phi$ that provides a positive key rate in the presence of leakage light, with the range being narrowest for Model 1. We observed for both types of sources that the range of $\phi$ that yields positive key rate narrows as the channel distance is increased; this means that source preparation flaws, especially in the test state, become a greater problem at further distances, unlike in the case of no leakage light where there is greater stability of the key rate with respect to $\phi$.

The main point of these simulations is to demonstrate that while the choice of test state is not so important when the source is ideal without side-channels, in the presence of leakage light, we must be careful to choose a test state that provides both good constraints on the encoded mode and on the leakage mode. While $\phi=\pi/2$ seemed to be the best choice for these models---coinciding with the BB84 states---we also observed cases when other values of $\phi$ produced the maximum key rate at a given distance. In a typical protocol, Alice and Bob simply choose the BB84 states and optimize the decoy state intensities as a function of distance; here, we see that in the presence of leakage light, there is additional benefit to optimizing over the polarization of the test states sent.

\bibliographystyle{apsrev}
\bibliography{main.bbl}

\end{document}